%% file: 12psc_arxiv.tex
\documentclass[twocolumn]{aastex62}
\UseRawInputEncoding

\newcommand{\Mjup}{\mbox{$M_\mathrm{Jup}$}}
\newcommand{\Msun}{\mbox{$M_{\odot}$}}

\usepackage{color}

\shorttitle{White Dwarf Companions Accelerating 12 Psc and HD 159062}
\shortauthors{Bowler et al.}
\begin{document}

\title{The McDonald Accelerating Stars Survey (MASS): \\
White Dwarf Companions Accelerating the Sun-like Stars 12 Psc and HD 159062}

\correspondingauthor{Brendan P. Bowler}
\email{bpbowler@astro.as.utexas.edu}

\author[0000-0003-2649-2288]{Brendan P. Bowler}
\affiliation{Department of Astronomy, The University of Texas at Austin, Austin, TX 78712, USA}

\author{William D. Cochran}
\affiliation{Center for Planetary Systems Habitability and McDonald Observatory, The University of Texas at Austin, Austin, TX 78712, USA}

\author{Michael Endl}
\affiliation{McDonald Observatory and the Department of Astronomy, The University of Texas at Austin, Austin, TX 78712, USA}

\author{Kyle Franson}
\affiliation{Department of Astronomy, The University of Texas at Austin, Austin, TX 78712, USA}

\author{Timothy D. Brandt}
\affiliation{Department of Physics, University of California, Santa Barbara, Santa Barbara, CA 93106, USA}

\author{Trent J. Dupuy}
\affiliation{Gemini Observatory, Northern Operations Center, 670 N. AÕohoku Place, Hilo, HI 96720, USA}

\author{Phillip J. MacQueen}
\affiliation{McDonald Observatory and the Department of Astronomy, The University of Texas at Austin, Austin, TX 78712, USA}

\author{Kaitlin M. Kratter}
\affiliation{Department of Astronomy, University of Arizona, Tucson, AZ 85721, USA}

\author{Dimitri Mawet}
\affiliation{Department of Astronomy, California Institute of Technology, Pasadena, CA 91125, USA}

\author{Garreth Ruane}
\affiliation{Department of Astronomy, California Institute of Technology, Pasadena, CA 91125, USA}

\begin{abstract}

We present the discovery of a white dwarf companion to the G1 V star 12 Psc 
found as part of a Keck adaptive optics imaging survey of long-term accelerating stars 
from the McDonald Observatory Planet Search Program.  
Twenty years of precise radial-velocity monitoring of 12 Psc with the Tull Spectrograph at the 
Harlan J. Smith telescope reveals a moderate radial acceleration ($\approx$10 m s$^{-1}$ yr $^{-1}$), 
which together with relative astrometry from Keck/NIRC2 and the astrometric acceleration 
between $Hipparcos$ and $Gaia$ DR2 yields a dynamical mass of $M_B$ = 0.605$^{+0.021}_{-0.022}$ \Msun \ for 12 Psc B,
a semi-major axis of 40$^{+2}_{-4}$~AU, and an eccentricity of 0.84$\pm$0.08.
We also report an updated orbit fit of the white dwarf companion to the metal-poor (but barium-rich) G9 V dwarf HD 159062
based on new radial velocity observations from the High-Resolution Spectrograph at the Hobby-Eberly Telescope and astrometry from Keck/NIRC2.
A joint fit of the available relative astrometry, radial velocities, and tangential astrometric acceleration 
yields a dynamical mass of $M_B$ = 0.609$^{+0.010}_{-0.011}$ \Msun \ for HD 159062 B,
a semi-major axis of 60$^{+5}_{-7}$~AU, and preference for circular orbits ($e$$<$0.42 at 95\% confidence).
12 Psc B and HD 159062 B join a small  list of resolved ``Sirius-like'' benchmark white dwarfs 
with precise dynamical mass measurements
which serve as valuable tests of white dwarf mass-radius cooling models and 
probes of AGB wind accretion onto their main-sequence companions.  

\end{abstract}


\section{Introduction} \label{sec:intro}

Dynamical masses represent anchor points of stellar astronomy.  
Direct mass measurements are important to calibrate models of stellar and substellar evolution,
especially during phases in which physical properties change significantly with time---for example
throughout the pre-main sequence; along the evolved subgiant and giant branches; and as white dwarfs,
brown dwarfs, and giant planets cool and fade over time (e.g., \citealt{Hillenbrand:2004bi}; \citealt{Simon:2019dh}; 
\citealt{Konopacky:2010kr}; \citealt{Bond:2017cx}; \citealt{Parsons:2017kt}; \citealt{Dupuy:2017ke}; 
\citealt{Snellen:2018aa}; \citealt{Brandt:2019ey}).
Masses are traditionally determined with absolute astrometry of visual binaries or
radial-velocity (RV) monitoring of either eclipsing or visual binaries.
Other approaches 
include modeling Keplerian rotation of resolved protoplanetary disks,
gravitational lensing, 
and transit-timing variations in the case of close-in planets (see \citealt{Serenelli:2020aa} for a recent review).

It is especially challenging to measure dynamical masses of non-transiting binaries 
when one component is faint, as is the case of white dwarf, brown dwarf,
or giant planet companions to stars.
With high-contrast adaptive optics (AO) imaging, there is a pragmatic trade-off between separation 
and contrast: short period companions reveal their orbits on faster timescales but are more challenging to detect,
whereas more distant companions are easier to image but orbit more slowly.  
Similarly, RV precision, stellar activity, and time baseline of the observations compete when 
measuring a radial acceleration induced on the star by the companion.
The optimal region in which radial reflex accelerations can be measured and faint 
companions
can be imaged with current facilities is $\sim$5-100 AU.
Most of the known benchmark white dwarf, brown dwarf, and giant planet companions 
fall in this range of orbital distances (see, e.g., Table 2 of \citealt{Bowler:2016jk}).

One of the most successful means of identifying these faint ``degenerate'' companions 
(whose pressure support predominantly originates from electron degeneracy) with
direct imaging has been by using radial accelerations on their host stars, which
can act as ``dynamical beacons'' that betray the presence of a distant companion.
Long-baseline RV surveys are especially well-suited for this task,
such as the
California Planet Survey (\citealt{Howard:2010dia}), 
Lick-Carnegie Exoplanet Survey (\citealt{Butler:2017km}), 
McDonald Observatory Planet Search (\citealt{Cochran:1993va}), 
Lick Planet Search (\citealt{Fischer:2014ew}), 
Anglo-Australian Planet Search (\citealt{Tinney:2001ha}), 
and CORALIE survey for extrasolar planets (\citealt{Queloz:2000aa}).
With baselines spanning several decades and sample sizes of thousands of targets, these programs
have facilitated the discovery and characterization of a growing list of substellar companions
(HR 7672 B, \citealt{Liu:2002fx}; 
HD 19467 B, \citealt{Crepp:2014ce}; 
HD 4747 B, \citealt{Crepp:2016fg}; 
HD 4113 C, \citealt{Cheetham:2018haa}; 
Gl 758 B, \citealt{Thalmann:2009ca}, \citealt{Bowler:2018gy}; 
HD 13724 B, \citealt{Rickman:2020aa};
HD 72946 B, \citealt{Maire:2020iu};
HD 19467 B,  \citealt{Maire:2020iu}; 
Gl 229 B, \citealt{Nakajima:1995bb}; \citealt{Brandt:2019kp})
and white dwarf companions 
(Gl 86 B, \citealt{Els:2001hd}, \citealt{Mugrauer:2005kt}; 
HD 8049 B, \citealt{Zurlo:2013kb}; 
HD 114174 B, \citealt{Crepp:2013ij}; 
HD 11112 B, \citealt{Rodigas:2016kc}; 
HD 169889 B, \citealt{Crepp:2018kf}; 
HD 159062 B, \citealt{Hirsch:2019cp})
with direct imaging.
Only a handful of these degenerate companions have dynamically measured masses, although
recent efforts to determine 
astrometric accelerations on their host stars 
using \emph{Hipparcos} and \emph{Gaia} are increasing these numbers (e.g., \citealt{Calissendorff:2018ee}; \citealt{Brandt:2019ey}; \citealt{Dupuy:2019cy}).

To find new benchmark companions and measure their dynamical masses,
we launched the McDonald Accelerating Stars Survey (MASS), a
high-contrast imaging program targeting stars with radial accelerations based on RV planet search programs
at McDonald Observatory.
The McDonald Observatory Planet Search began in 1987 at the 2.7-m Harlan J. Smith Telescope 
and is among the oldest radial velocity planet surveys (\citealt{Cochran:1993va}). The most recent phase 
of the survey using the Tull Spectrograph commenced in 1998 and continues today.
In addition to discoveries of giant planets spanning orbital periods of a few days to over ten years 
(e.g., \citealt{Cochran:1997ta}; \citealt{Hatzes:2003fe}; \citealt{Robertson:2012hc}; \citealt{Endl:2016kk}), 
many shallow long-term accelerations have been identified over the past three decades. 
Accelerating stars in our sample also draw from a planet search around 145 metal-poor stars 
using the 9.2-m Hobby-Eberly Telescope's High-Resolution Spectrograph (HRS).  This program operated from 2008 to 2013
and, like the McDonald Observatory Planet Search, identified both planets and longer-term 
radial accelerations (\citealt{Cochran:2008cg}). 

In \citet{Bowler:2018gy} we presented an updated orbit and mass measurement of the late-T dwarf Gl 758 B
as part of this program based on new imaging data and RVs from McDonald Observatory, Keck Observatory, and the Automated Planet Finder.
The mass of Gl 758 B was subsequently refined in \citet{Brandt:2019ey} by 
taking into account the proper motion difference between \emph{Hipparcos} and \emph{Gaia}.
Here we present the discovery and dynamical mass measurement of a faint white dwarf companion
to the Sun-like star 12 Psc based on a long-term RV trend of its host star
from the McDonald Observatory Planet Search.
In addition, we present 
an updated orbit and mass measurement of HD 159062 B, a white dwarf companion to an accelerating
G9 V star recently discovered by \citet{Hirsch:2019cp}
and independently identified in our program using radial velocities from HRS.
These objects join only a handful of other resolved white dwarf companions with dynamical mass measurements.

This paper is organized as follows.  In Section~\ref{sec:overview} we provide an overview of the properties of
12 Psc and HD 159062.
Section \ref{sec:obs} describes the RV and imaging observations of these systems from McDonald Observatory
and  Keck Observatory.  The orbit fits and dynamical mass measurements for both companions are detailed in 
Section~\ref{sec:results}.  Finally, we discuss the implications of the mass measurements for the 
evolutionary history of the system in Section~\ref{sec:discussion}.

\section{Overview of 12 Psc and HD 159062}{\label{sec:overview}}

12 Psc  (=HD 221146, HIP 115951) is a bright ($V$ = 6.9 mag) G1 V dwarf (\citealt{Gray:2006ca}) located at a 
parallactic distance of 36.2 $\pm$ 0.06 pc  (\citealt{GaiaCollaboration:2018io}).
Spectroscopy and isochrone fitting imply a slightly more massive, older, and more metal rich analog to the Sun.
For example, \citet{Soto:2018bl} find an age of 5.3$^{+1.1}_{-1.0}$ Gyr, a mass of 1.11 $\pm$ 0.05~\Msun,
a metallicity of [Fe/H] = 0.13 $\pm$ 0.10 dex, and an effective temperature of 5950 $\pm$ 50 K.
This is in good agreement with other recent analysis from 
\citet{Tsantaki:2013dc}, \citet{Marsden:2014bd}, and \citet{AguileraGomez:2018bn}.
The old age is bolstered by the low activity level, with $\log R'_\mathrm{HK}$ values 
ranging from --5.06 dex to --4.86 dex (e.g., \citealt{Isaacson:2010gk}; \citealt{Murgas:2013hh}; \citealt{Saikia:2018dh}).
A summary of the physical, photometric, and kinematic properties of 12~Psc can be found in Table~\ref{tab:12pschostprop}.

HD 159062 is an old, metal-poor, main-sequence G9 V star (\citealt{Gray:2003fza}) located at a distance of 21.7 pc (\citealt{GaiaCollaboration:2018io}).
\citet{Hirsch:2019cp} derive a mass of 0.76 $\pm$ 0.03~\Msun \ using spectroscopically-derived physical properties
and stellar isochrones.
A wide range of ages have been determined for HD 159062 in the literature: \citet{Isaacson:2010gk}  and \citet{Hirsch:2019cp} 
find activity-based ages of $\approx$6~Gyr and $\approx$7~Gyr using $R'_\mathrm{HK}$ values, while
typical isochrone-based ages range from 9.2 $\pm$ 3.5 Gyr from \citet{Luck:2017jd} to 13.0$^{+1.4}_{-2.4}$ Gyr from \citet{Brewer:2016gf}.
\citet{Brewer:2006wq} measure a metallicity of [Fe/H] = --0.50 dex and find that HD 159062 has an 88\% probability of belonging
to the thick disk based on its kinematics.  HD 159062 also exhibits an enhancement of $\alpha$-capture 
elements such as [Mg/Fe], [Si/Fe], and [Ca/Fe], further supporting membership in the thick disk.  
The low metallicity, enhanced [$\alpha$/Fe] abundances, 
and thick-disk kinematics point to an older age for this system.
\citet{Brewer:2006wq} also note that HD 159062 exhibits substantially enhanced $s$-process elements and
suggest this could have been caused by mass transfer from an evolved AGB companion.
This scenario is supported by \citet{Fuhrmann:2017aa}, who find anomalously high barium abundance and conclude 
that HD 159062 may harbor a white dwarf companion.
This prediction was realized with the discovery of HD 159062 B by \citet{Hirsch:2019cp}
using a long-baseline RV trend from Keck/HIRES and follow-up adaptive optics imaging.
They determine a dynamical mass of 0.65$^{+0.12}_{-0.04}$~\Msun \ for HD 159062 B,
which was refined by Brandt et al. (submitted) to 0.617$^{+0.013}_{-0.012}$~\Msun \ after adding in the 
astrometric acceleration induced on the host star using \emph{Hipparcos} and \emph{Gaia} (see \ref{sec:hgca}).

\begin{deluxetable}{lcc}
\renewcommand\arraystretch{0.9}
\tabletypesize{\small}
\setlength{ \tabcolsep } {.1cm} 
\tablewidth{0pt}
\tablecolumns{3}
\tablecaption{Properties of 12 Psc\label{tab:12pschostprop}}
\tablehead{
       \colhead{Parameter} & \colhead{Value}  & \colhead{Reference}  
        }   
\startdata
\cutinhead{Physical Properties}
$\alpha_{2000.0}$  &  23:29:30.31  &  $\cdots$ \\
$\delta_{2000.0}$  &  --01:02:09.1  &  $\cdots$ \\
$\pi$ (mas) & 27.60 $\pm$ 0.05  &  1 \\
Distance (pc) & 36.23 $\pm$ 0.06 &  1 \\
SpT  &  G1 V  &  2 \\
Mass (\Msun)  &  1.11 $\pm$ 0.05  &  3 \\
Age (Gyr)   &  5.3 $\pm$ 1.1 & 3 \\
$T_\mathrm{eff}$ (K)  &  5950 $\pm$ 50  & 3  \\
$\log (L_\mathrm{bol}$/$L_{\odot}$)  &  0.358 $\pm$ 0.08  &  4  \\
$\log g$ (dex) [cgs]  &  4.34 $\pm$ 0.3  &  3  \\
$R$ (R$_{\odot}$)  &  1.32 $\pm$ 0.03  &  3  \\
$\rm{[Fe/H]}$ (dex)  &  +0.13 $\pm$ 0.10  &  3  \\
$v \sin i$ (km s$^{-1}$)  &  2.3 $\pm$ 0.2 & 3  \\
$\log R'_\mathrm{HK}$  &  --5.07 $\pm$  0.01  &  4 \\
Proj. Sep. ($''$)  &  1.6  &  5  \\
Proj. Sep. (AU)  &  58  &  5  \\
$dv_r$/$dt$ (m s$^{-1}$ yr$^{-1}$) &  10.60 $\pm$ 0.13  &  5  \\
\cutinhead{Photometry}
$V$ (mag)  &  6.92 $\pm$ 0.04  &  6 \\
$Gaia$ $G$ (mag) & 6.7203 $\pm$ 0.0003  &  1 \\
$J$ (mag)  &  5.77 $\pm$ 0.01  &  7 \\
$H$ (mag)  &  5.49 $\pm$ 0.03  & 7 \\
$K_S$ (mag)  &  5.40 $\pm$ 0.01  & 7 \\
\cutinhead{HGCA Kinematics\tablenotemark{a}}
$\mu_{\alpha, \mathrm{Hip}}$ (mas yr$^{-1}$) &  --11.97 $\pm$ 1.09  &  8 \\
$\mu_{\alpha, \mathrm{Hip}}$ Epoch (yr) & 1991.348 & 8 \\
$\mu_{\delta, \mathrm{Hip}}$ (mas yr$^{-1}$) &  --28.50 $\pm$ 0.84  &  8 \\
$\mu_{\delta, \mathrm{Hip}}$ Epoch (yr) & 1991.277 & 8 \\
$\mu_{\alpha, \mathrm{HG}}$ (mas yr$^{-1}$) &  --11.32 $\pm$ 0.03  &  8 \\
$\mu_{\delta, \mathrm{HG}}$ (mas yr$^{-1}$) &  --25.66 $\pm$ 0.02  &  8 \\
$\mu_{\alpha, \mathrm{Gaia}}$ (mas yr$^{-1}$) &  --10.54 $\pm$ 0.12  &  8 \\
$\mu_{\alpha, \mathrm{Gaia}}$ Epoch (yr) & 2015.563 & 8 \\
$\mu_{\delta, \mathrm{Gaia}}$ (mas yr$^{-1}$) &  --24.24 $\pm$ 0.09  &  8 \\
$\mu_{\delta, \mathrm{Hip}}$ Epoch (yr) & 2015.669 & 8 \\
$\Delta \mu_{\alpha, \mathrm{Gaia-HG}}$ (mas yr$^{-1}$) & 0.78 $\pm$ 0.12 & 8 \\
$\Delta \mu_{\delta, \mathrm{Gaia-HG}}$ (mas yr$^{-1}$) & 1.42 $\pm$ 0.09 & 8 \\
$d \mu_{\alpha \delta}/dt$ (m s$^{-1}$ yr$^{-1}$) & 22.9 $\pm$ 1.4 &  8 \\
\enddata
\tablerefs{(1) \citet{GaiaCollaboration:2018io}; (2) \citet{Gray:2006ca}; (3) \citet{Soto:2018bl}; 
(4) \citet{Marsden:2014bd}; (5) This work; (6) \citet{Richmond:2000aa}; (7) \citet{Cutri:2003tp}; (8) \citet{Brandt:2018dja}.}
\tablenotetext{a}{\emph{Hipparcos}-\emph{Gaia} Catalog of Accelerations (\citealt{Brandt:2018dja}).  Proper motions in R.A. include a factor of $\cos \delta$.}
\end{deluxetable}

\section{Observations}{\label{sec:obs}}

% Figure 1

\begin{figure}
  \vskip -0.2 in
  \hskip -0.8 in
  \resizebox{5in}{!}{\includegraphics{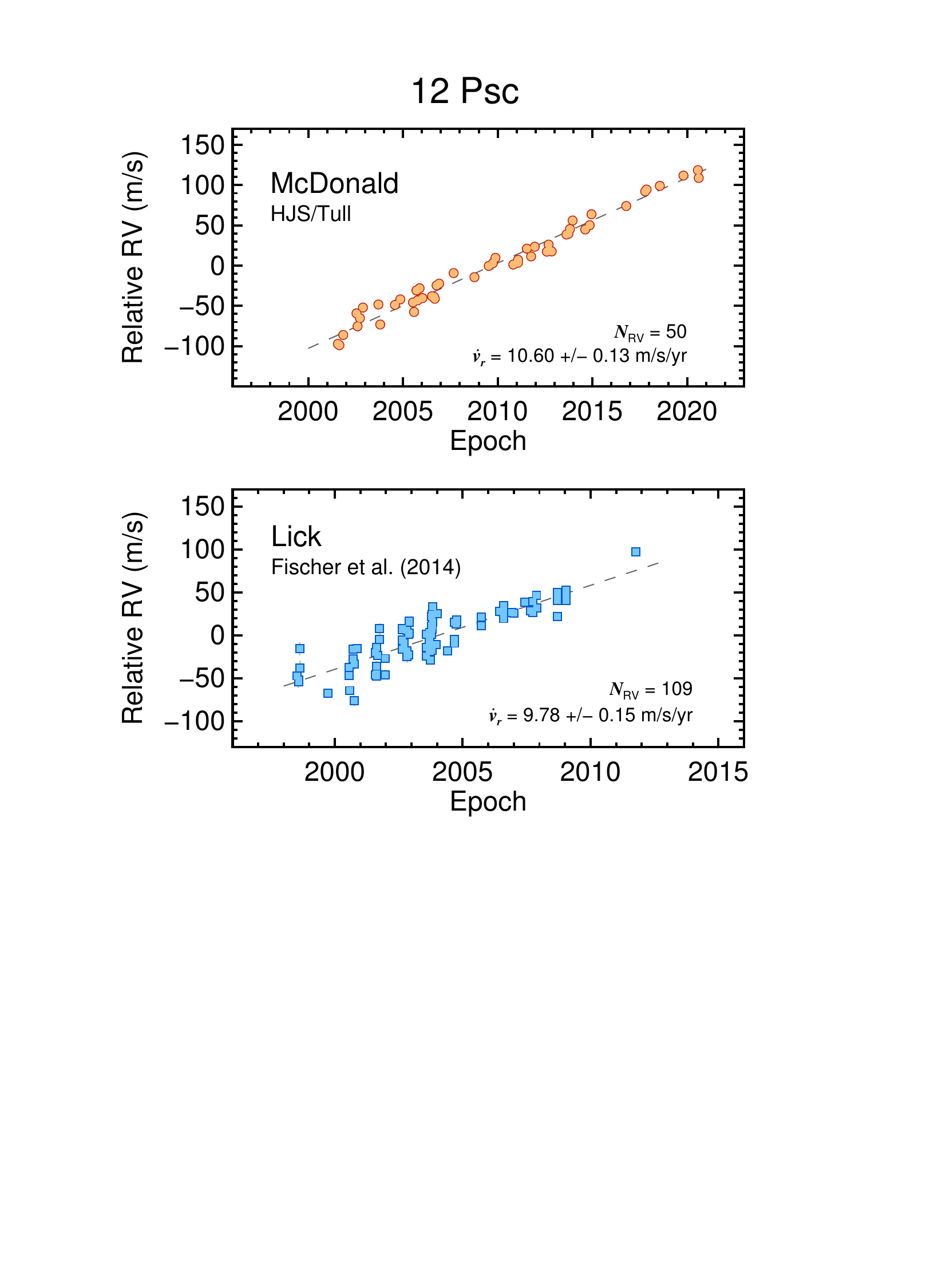}}
  \vskip -2.5 in
  \caption{Radial velocities of 12 Psc from McDonald Observatory (top) and 
  Lick Observatory (bottom; \citealt{Fischer:2014ew}) spanning a total baseline of over twenty years (1998 to 2020).  
  A strong radial acceleration is evident in both datasets.  A linear fit gives
  a slope of 10.60 $\pm$ 0.13 m s$^{-1}$ yr$^{-1}$ for the Tull RVs and
  9.78 $\pm$ 0.15 m s$^{-1}$ yr$^{-1}$ for the Lick RVs.   \label{fig:12psc_rvs} } 
\end{figure}

\subsection{Radial Velocities} \label{sec:rvs}

\subsubsection{Harlan J. Smith Telescope/Tull Spectrograph Radial Velocities of 12 Psc}
50 RV measurements of 12 Psc were obtained with the Tull Coud\'{e} spectrograph (\citealt{Tull:1995tn})
at McDonald Observatory's 2.7-m Harlan J. Smith telescope between 2001 and 2020.   
All observations used the 1$\farcs$2 slit, resulting in a resolving power of $R$ $\equiv$ $\lambda$/$\Delta \lambda$ $\approx$ 60,000.
A temperature-stabilized gas cell containing molecular iodine vapor (I$_2$) is mounted in the light path before the slit entrance,
enabling precise RV measurements 
with respect to an iodine-free template following the description in \citet{Endl:2000ui}.  
RVs are subsequently corrected for Earth's barycentric motion as well as the small secular acceleration for 12 Psc (0.000582 m s$^{-1}$ yr$^{-1}$).
The time of observation is corrected to the barycentric dynamical time as observed at the solar
system barycenter.
Observations starting in 2009 take into account the flux-weighted 
barycentric correction 
of each observation using an exposure meter.  
The RVs are shown in Figure~\ref{fig:12psc_rvs} and are listed in Table~\ref{tab:12psc_rvs}\footnote{Note that 
the measured RVs are determined with respect to a dense set of iodine absorption lines
rather than an absolute reference, such as a stable RV standard.  The zero point is therefore arbitrary.}.
The median measurement uncertainty is 5.1 m s$^{-1}$.

12 Psc shows a constant acceleration away from the Sun with no obvious signs of curvature, 
indicating that a companion orbits this star with a period substantially longer than 
the time baseline of the observations ($P$$\gg$20~yr).  
A linear fit to the McDonald RVs gives a radial acceleration of $dv_r$/$d t$ = 10.60 $\pm$ 0.13 m s$^{-1}$ yr$^{-1}$.

12 Psc was also observed with the Hamilton Spectrograph at Lick Observatory 
as part of the Lick Planet Search (\citealt{Fischer:2014ew}) between 1998 and 2012.  
109 RVs were obtained with a typical precision of 4.4 m s$^{-1}$.  A linear fit to the Lick RVs gives
a slope of $dv_r$/$d t$ = 9.78 $\pm$ 0.15 m s$^{-1}$ yr$^{-1}$ (Figure~\ref{fig:12psc_rvs}).  
This is slightly shallower than the slope from the McDonald RVs (at the 4$\sigma$ level),
which may indicate modest change in acceleration between the mean epochs of both datasets 
(2004.0 for Lick and 2009.7 for McDonald).

\input{12psc_rvs_tull_sample.tex}

\input{hd159062_rvs_tull_sample.tex}

% Figure 2

\begin{figure}
  \vskip -0.2 in
  \hskip -0.8 in
  \resizebox{5in}{!}{\includegraphics{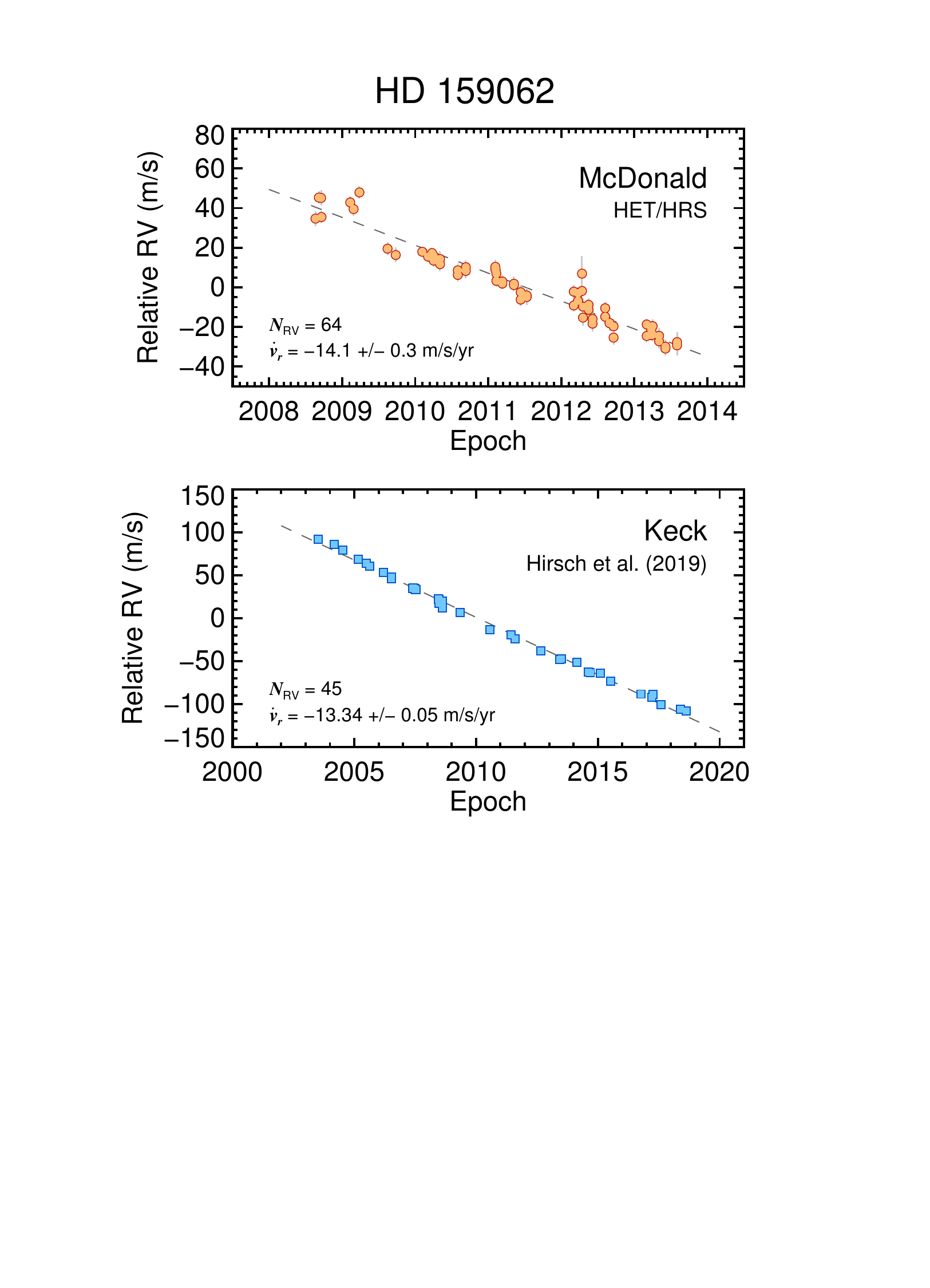}}
  \vskip -2.5 in
  \caption{Radial velocities of HD 159062 from HET/HRS (top) 
  and Keck/HIRES (bottom; \citealt{Hirsch:2019cp}).
  A linear fit to the HRS and HIRES datasets give accelerations of 
  --14.1 $\pm$ 0.3 m s$^{-1}$ yr$^{-1}$ and --13.34 $\pm$ 0.05 m s$^{-1}$ yr$^{-1}$, respectively.
  \label{fig:hd159062_rvs} } 
\end{figure}

\subsubsection{Hobby-Eberly Telescope/High-Resolution Spectrograph Radial Velocities of HD 159062}

HD 159062 was monitored with  HRS at the Hobby-Eberly Telescope
between 2008 and 2014.  HRS is a 
fiber-fed echelle spectrograph located in the basement of the HET (\citealt{Tull:1998jy}),
and is passively (rather than actively) thermally and mechanically stabilized.
A temperature controlled I$_2$ cell is mounted in front of the entrance slit
and is used as a reference for RV measurements.
64 high-SNR spectra were acquired with the HET's 
flexible queue scheduling system (\citealt{Shetrone:2007fu})
with a resolving power of $R$$\approx$60,000 between  4110~\AA \ and 7875~\AA.
Relative RVs are measured in spectral chunks following the procedure
described in \citet{Cochran:2003ua} and \citet{Cochran:2004kh}.
The median RV uncertainty is 3.4 m s$^{-1}$.

The HRS RVs of HD 159062 are shown in Figure~\ref{fig:hd159062_rvs}.  
We find an acceleration of $dv_r$/$d t$ = --14.1 $\pm$ 0.3 m s$^{-1}$ yr$^{-1}$,
which is similar to the slope of --13.30 $\pm$ 0.12 m s$^{-1}$ yr$^{-1}$
from \citet{Hirsch:2019cp} based on 45 Keck/HIRES RVs spanning 2003 to 2019.
A slight curvature is seen in the HIRES data; this changing acceleration is not evident in our HRS data, most likely 
owing to the shorter time baseline compared to the HIRES data.
A list of our HRS RVs can be found in Table~\ref{tab:hd159062_rvs}.

\subsection{Keck/NIRC2 Adaptive Optics Imaging} \label{sec:nirc2}

We imaged 12 Psc and HD 159062 with the 
NIRC2 infrared camera in its narrow configuration (9.971~mas pix$^{-1}$ plate scale; \citealt{Service:2016gk})  
using natural guide star adaptive optics at 
Keck Observatory (\citealt{Wizinowich:2000hl}; \citealt{Wizinowich:2013dz}).
12 Psc was initially targeted as part of this program on 2017 October 10 UT with subsequent observations
on 2018 December 24 UT and 2019 July 07 UT.  HD 159062 was observed on 2017 October 10 UT
and 2019 July 07 UT.  
For each observation, the star was centered behind the partly transparent 600 mas diameter coronagraph
to avoid saturation when reading out the full 10$\farcs$2$\times$10$\farcs$2 array.  
Most sequences consist of five coronagraphic (``reconnaissance'') images with the $H$- or $K_S$-band filters to search 
for readily identifiable long-period stellar or substellar companions.

12 Psc B and HD 159062 B were immediately evident in the raw frames, although it was not clear whether 
they were faint white dwarfs, brown dwarfs, or background stars at the time of discovery.  We 
also acquired deeper sequences in pupil-tracking mode (or angular differential imaging; \citealt{Marois:2006df}) in 
July 2019 to search for additional companions at smaller separations.  
These observations consisted of forty 30-second frames; for 12 Psc the total field 
rotation was 11.6$^{\degr}$ and for HD 159062 the total rotation was 16.4$^{\degr}$.
On the October 2017 and July 2019 nights we also acquired short unsaturated images immediately 
following the coronagraphic images to photometrically calibrate the deeper frames.
A summary of our observations can be found in Table~\ref{tab:newastrometry}.

We searched the Keck Observatory Archive and found that 
12 Psc was also imaged on two separate occasions with NIRC2 in 
September 2004 and July 2005 (PI: M. Liu) in $J$ and $K_p$ bands, respectively.
Both sequences consist of coronagraphic images similar to our observations with the host star
centered behind the occulting spot.
12 Psc B is visible in both frames, offering a 15-year astrometric baseline to test for common
proper motion and measure orbital motion.

% Figure 3

\begin{figure*}
  \vskip -1 in
  \hskip -0.2 in
  \resizebox{9.in}{!}{\includegraphics{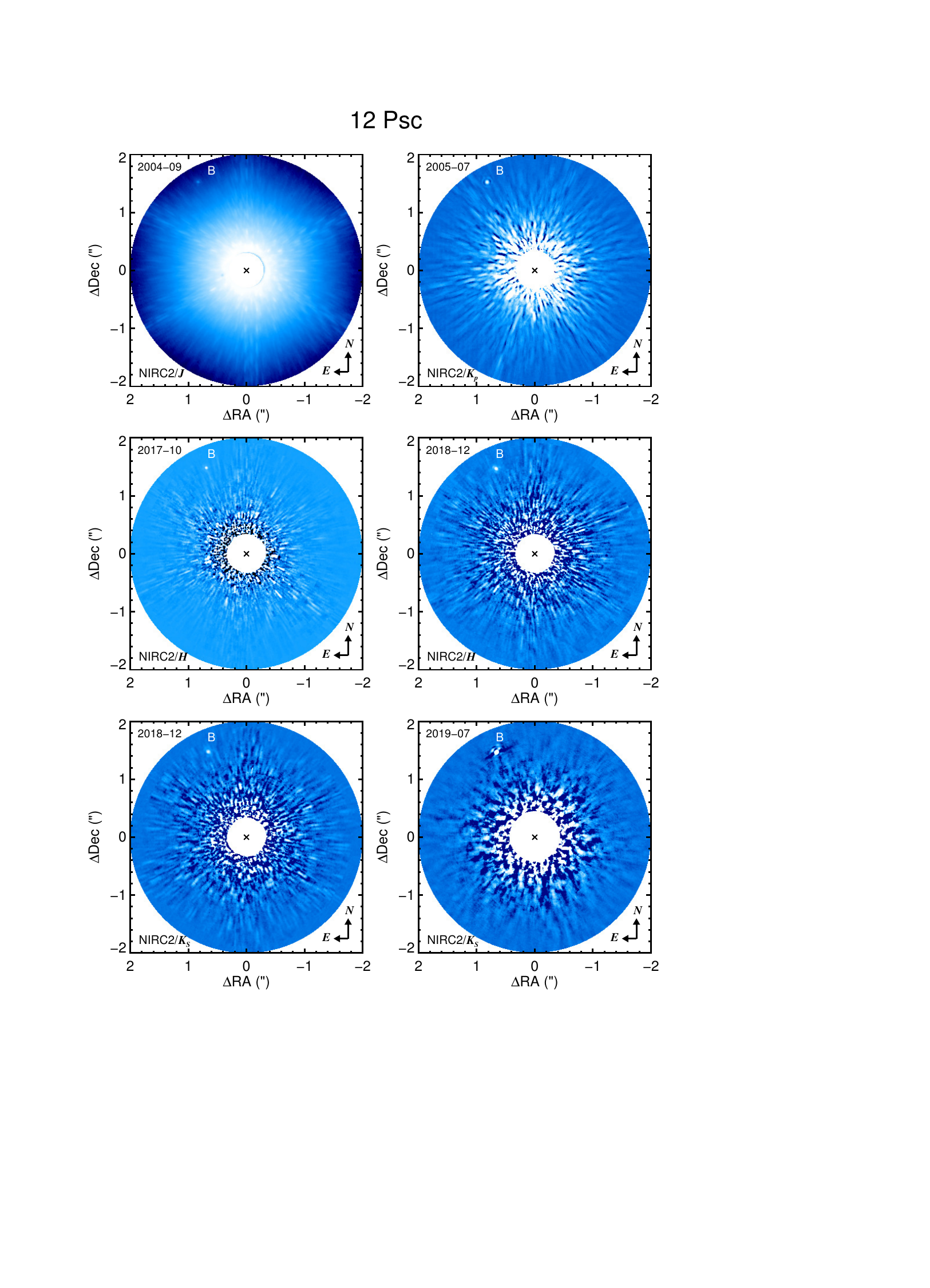}}
  \vskip -2.9 in
  \caption{NIRC2 adaptive optics images of 12 Psc B between September 2004 and July 2019.  
  PSF subtraction has been carried out for all observations
  except the 2004 $J$-band dataset (see Table~\ref{tab:newastrometry} for details).  The position of the host star 
  behind the 600~mas diameter coronagraph (masked out in these images) is 
  marked with an ``$\times$''.  North is up and East is to the left.  \label{fig:nirc2_12psc} } 
\end{figure*}

Basic data reduction was carried out in the same fashion for all images, which consisted of flat fielding using dome flats,
correction for bad pixels and cosmic rays, and correction for geometric field distortion using the solutions derived
by \citet{Yelda:2010ig} for observations taken before April 2015 (when the adaptive optics system 
was realigned) and \citet{Service:2016gk} for images taken after that date.
For both distortion solutions, the direction of celestial north was calibrated by tying NIRC2 observations
of globular clusters to distortion-corrected stellar positions obtained with the \emph{Hubble Space Telescope}.
The resulting precision in the north orientation is $\approx$0.001--0.002$\degr$---much less than
the typical measurement uncertainties for AO-based relative astrometry.

\begin{figure*}
  \vskip -1 in
  \hskip -0.5 in
  \resizebox{10.in}{!}{\includegraphics{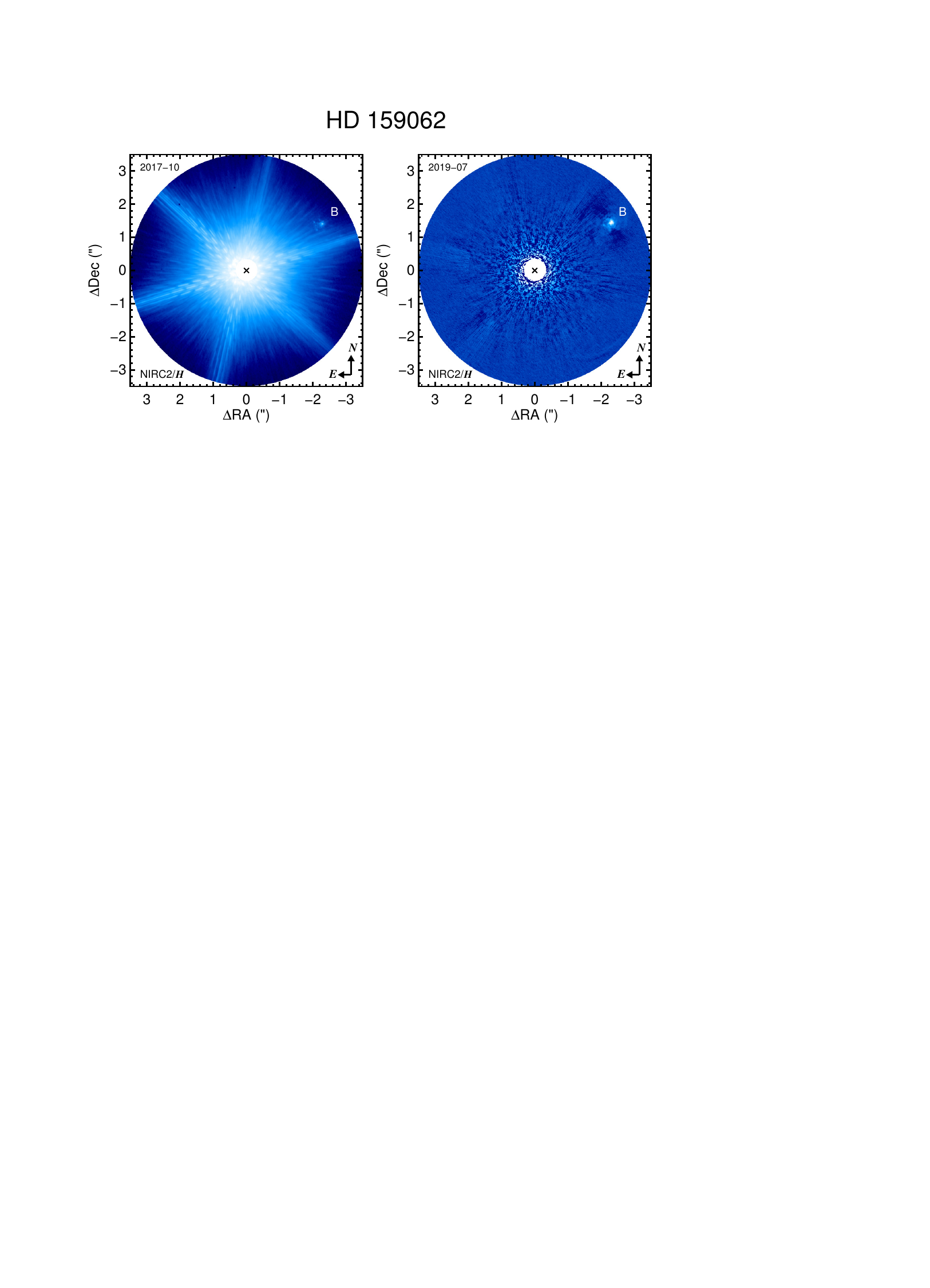}}
  \vskip -9.2 in
  \caption{NIRC2 adaptive optics $H$-band observations of HD 159062 and its white dwarf companion.  
  No additional point sources at smaller separations are evident 
  in the post-processed July 2019 ADI sequence. 
  \label{fig:nirc2_hd159062} } 
\end{figure*}

Images in each coronagraph sequence are aligned by fitting a 2D elliptical 
Gaussian to the host star, which is visible behind the partly transparent mask, then shifting each image
with sub-pixel precision to a common position.  
We attempted several approaches to PSF subtraction for each sequence in order to increase the SNR
of the companion: subtraction of a scaled median image of the sequence; a conservative implementation of the Locally Optimized
Combination of Images (LOCI; \citealt{Lafreniere:2007bg}); an aggressive form of LOCI with a more restrictive angular tolerance parameter; 
LOCI using 100 images selected in an automated fashion from a NIRC2 reference 
PSF library comprising $>$2$\times$10$^{3}$ registered coronagraph frames; and optional masking of the companion during
PSF subtraction.
Details of these methods and the NIRC2 PSF library are described in \citet{Bowler:2015ja} and \citet{Bowler:2015ch}.
We attempted PSF subtraction for the 2004 observations of 12 Psc and the 2017 observations of HD 159062 but 
strong systematics were present in the residuals.
The final processed images we adopt for this study are shown in Figures~\ref{fig:nirc2_12psc} and \ref{fig:nirc2_hd159062}.
Table~\ref{tab:newastrometry} lists the observations and the adopted PSF subtraction method.

In general, astrometry and relative photometry of point sources measured directly from processed (PSF-subtracted) images 
can be biased as a result of self-subtraction and non-uniform field of view rotation.
These effects can be especially severe for longer angular differential imaging datasets (e.g., \citealt{Marois:2010hs}).
When possible we use the negative PSF injection approach described in \citet{Bowler:2018gy} to mitigate these biases.
This entails adding a PSF with a negative amplitude close to the position of the point source in the raw images,
running PSF subtraction at that position, and measuring the RMS of the residuals in a circular aperture.  This process is then 
iteratively repeated by varying the astrometry ($\rho$ and $\theta$) at the sub-pixel level and the flux ratio (simply the amplitude 
of the negative PSF) using the \texttt{amoeba} algorithm (\citealt{Nelder:1965tk}; \citealt{Press:2007vx}) until the resulting RMS is minimized.  
This method requires a PSF model; if unsaturated frames of the host star are taken close in time to the ADI sequence
(to avoid changes in atmospheric conditions and AO correction), this approach can be used to reliably measure the contrast 
between the host star and the faint point source.  When no unsaturated frames are available, any PSF can be used to  measure 
astrometry (but not relative photometry).

We use this negative PSF injection approach to measure astrometry for the ADI datasets of 12 Psc and HD 159062 taken in July 2019.
In both cases, images of 12 Psc without the coronagraph mask are used as the 
PSF model since unsaturated images of HD 159062 were not taken 
on July 2019.  We also use this approach to measure relative photometry and astrometry of 12 Psc for the October 2017
dataset because unsaturated frames were acquired.  Uncertainties in the relative photometry are estimated using the mean
of the final ten iterations of the \texttt{amoeba} downhill simplex algorithm as the negative PSF separation, P.A., and
amplitude settle in at values that minimize the RMS at the position of the companion.
For all other observations, relative astrometry is directly computed from the unprocessed images.  
Total astrometric uncertainties are derived following \citet{Bowler:2018gy} and
incorporate random measurement uncertainties; systematic errors from the distortion solution; 
uncertainty in the plate scale and north angle; and shear (PSF blurring) caused by field rotation within an exposure.
Because the coronagraph mask may introduce an additional uncalibrated 
source of systematic uncertainty (e.g., \citealt{Konopacky:2016gv}; \citealt{Bowler:2018gy}),
we conservatively adopt 5 mas and 0.1$\degr$ as the floor for separation and position angle uncertainties,
respectively.
Final values for 12 Psc B and HD 159062 B can be found in Table~\ref{tab:newastrometry}.\footnote{Note that position
angles in this work correspond to the angle from celestial north through east at the epoch of observation, not for J2000.}

% Figure 4

\begin{deluxetable*}{lccccccc}
\renewcommand\arraystretch{0.9}
\tabletypesize{\small}
\setlength{ \tabcolsep }{.1cm} 
\tablewidth{0pt}
\tablecolumns{8}
\tablecaption{Keck/NIRC2 Adaptive Optics Imaging\label{tab:newastrometry}}
\tablehead{
       \colhead{UT Date}  & \colhead{Epoch} & \colhead{$N$$\times$Coadds$\times$$t_\mathrm{exp}$} & \colhead{Filter} &  \colhead{Sep.}  & \colhead{P.A.} &  \colhead{Contrast} & \colhead{PSF}  \\
       \colhead{(Y-M-D)}   & \colhead{(UT)}     & \colhead{(s)}                                                                           &  & \colhead{($''$)}  & \colhead{($\degr$)}  & \colhead{($\Delta$ mag)} &  \colhead{Sub.\tablenotemark{a}} 
        }   
\startdata
\cutinhead{12 Psc B}
2004 Sep 08  &  2004.688  &  7 $\times$ 1 $\times$ 30 &  $J$+cor600  &  1.723 $\pm$ 0.005  &  28.63 $\pm$ 0.10  & $\cdots$ & $\cdots$ \\ 
2005 Jul 15   &  2005.536  &  12 $\times$ 2 $\times$ 15 &  $K_p$+cor600  &  1.720 $\pm$ 0.005  &  28.5 $\pm$ 0.2  & $\cdots$ & 1 \\
2017 Oct 10   &  2017.773  &  5 $\times$ 6 $\times$ 5 &  $H$+cor600  &  1.623 $\pm$ 0.005  &  25.10 $\pm$ 0.14  & 10.38 $\pm$ 0.16  & 2, 3, 4 \\   
2018 Dec 24   &  2018.978  &  5 $\times$ 6 $\times$ 5 &  $H$+cor600  &  1.600 $\pm$ 0.005  &  24.33 $\pm$ 0.12  & $\cdots$  & 2, 3, 4 \\  
2018 Dec 24   &  2018.978  &  5 $\times$ 6 $\times$ 5 &  $K_S$+cor600  &  1.604 $\pm$ 0.005  &  24.26 $\pm$ 0.13  & $\cdots$  & 2, 3, 4 \\  
2019 Jul 07   &  2019.514   &   40 $\times$ 3 $\times$ 10 &  $K_S$+cor600  &  1.592 $\pm$ 0.005  &  24.4 $\pm$ 0.2  & 10.53 $\pm$ 0.01  & 2  \\ 
\cutinhead{HD 159062 B}
2017 Oct 10  &  2017.773   &   5 $\times$ 5 $\times$ 3 &  $H$+cor600  &  2.663 $\pm$ 0.005  &  301.34 $\pm$ 0.11  & $\cdots$  & $\cdots$ \\ 
2019 Jul 07  &  2019.513   &   40 $\times$ 10 $\times$ 3 &  $H$+cor600  &  2.702 $\pm$ 0.005  &  301.9 $\pm$ 0.4  & $\cdots$  & 2, 3  \\
\enddata
\tablenotetext{a}{PSF subtraction method: (1) scaled median subtraction;   (2) ``conservative LOCI'' with parameters 
$W$ = 5, $N_A$ = 300, $g$ = 1, $N_{\delta}$ = 1.5, and $dr$ = 2;   
(3) 100 additional images used from PSF reference library; (4) companion masked during PSF subtraction.
See \citet{Bowler:2015ja} and \citet{Bowler:2015ch} for additional details.} 
\end{deluxetable*}

\section{Results}{\label{sec:results}}

\subsection{Common Proper Motion}{\label{sec:cpm}}

\begin{figure}
  \vskip -0.2 in
  \hskip -0.2 in
  \resizebox{4.7in}{!}{\includegraphics{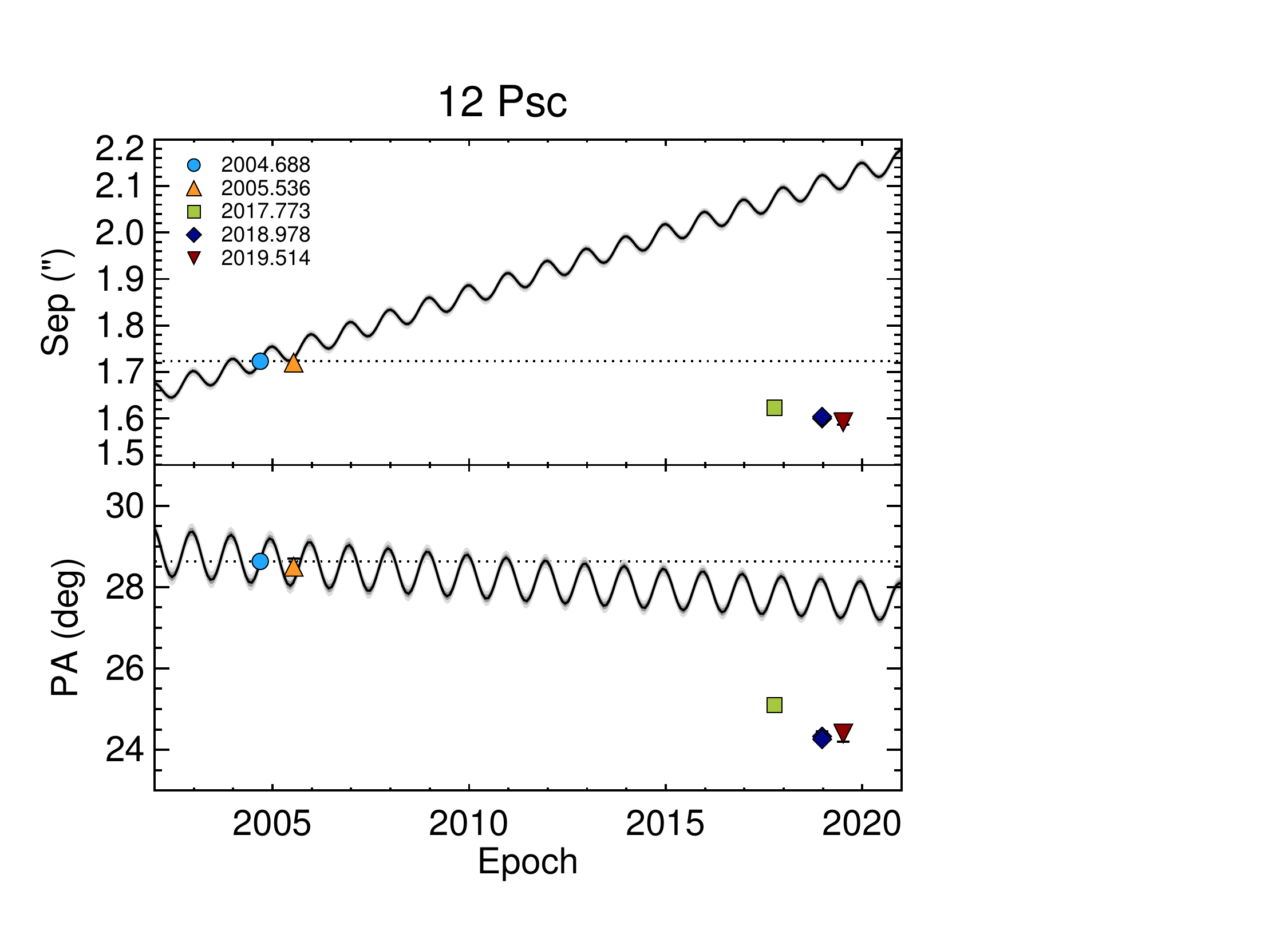}}
  \vskip -.1 in
  \caption{Relative astrometry of 12 Psc B.  Based on the first imaging epoch in 2004, the separation (top panel)
  and position angle (bottom panel) of a stationary source would follow the background track
  shown in black as a result of proper and parallactic motion of the host star.  12 Psc B is clearly comoving
  and shows significant orbital motion. \label{fig:background_12psc} } 
\end{figure}

Figure~\ref{fig:background_12psc} shows the expected relative motion of a stationary background source 
due to the proper and projected parallactic motion of 12 Psc.
12 Psc B is clearly comoving and exhibits significant orbital motion in separation and P.A.
A linear fit of the astrometry as a function of time gives a slope of --8.6 mas yr$^{-1}$ in separation
and --0.30$\degr$ yr$^{-1}$  in P.A.

\begin{figure}
  \vskip -0.2 in
  \hskip -0.2 in
  \resizebox{4.7in}{!}{\includegraphics{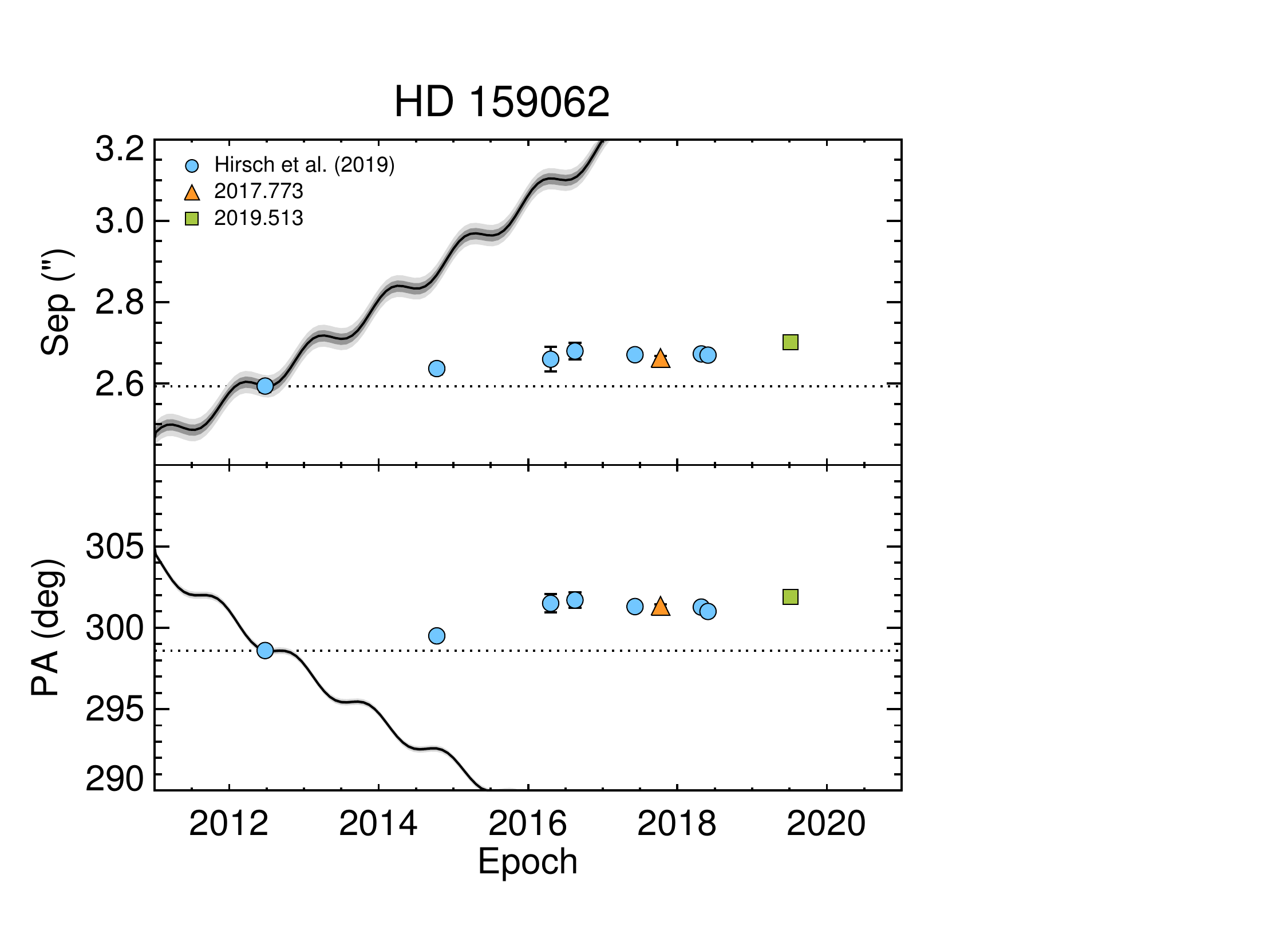}}
  \vskip -.1 in
  \caption{Same as Figure~\ref{fig:background_12psc} but for the white dwarf HD 159062 B.
  Blue circles show astrometry from \citet{Hirsch:2019cp}.  Our new observations
  in 2017 (triangle) and 2019 (square) follow a similar trend of increasing separation and P.A. over time.
  \label{fig:background_hd159062} } 
\end{figure}

Relative astrometry of HD 159062 B from \citet{Hirsch:2019cp} and our new observations are shown in 
Figure~\ref{fig:background_hd159062}.  HD 159062 B is moving away from its host at a rate of 13.8~mas yr$^{-1}$.
The rate of change in P.A. is 0.47$\degr$ yr$^{-1}$.  

% Figure 5

% Figure 6

\subsection{The Nature of 12 Psc B}

Based solely on its brightness ($H$=15.87~$\pm$~0.16 mag; $K_S$ = 15.93 $\pm$ 0.2~mag; 
$M_H$ = 13.08 $\pm$ 0.16~mag; $M_{K_S}$=13.1 $\pm$ 0.2~mag),
12 Psc B could  be either a brown dwarf or a white dwarf companion.
If it is a brown dwarf, its absolute magnitude would imply a spectral type near the L/T transition
and an $H$--$K_S$ color of $\approx$0.6~mag (\citealt{Dupuy:2012bp}).  
The measured color of 12 Psc B ($H$--$K_S$=--0.1~$\pm$~0.3~mag)
is significantly bluer than this, although the photometric uncertainties are large.
To estimate the expected mass of the companion assuming 12 Psc B is a brown dwarf,
we use a $K_S$-band bolometric correction from \citet{Filippazzo:2015dv} to infer a
luminosity of log~$L_\mathrm{bol}/L_{\odot}$=--4.61 $\pm$ 0.10~dex.  Based on the age
of the host star (5.3~$\pm$ 1.1~Gyr; \citealt{Soto:2018bl}), substellar evolutionary models
imply a mass near the hydrogen burning limit (77.8 $\pm$ 1.7~\Mjup \ using the 
\citealt{Burrows:1997jq} models and 69 $\pm$ 3~\Mjup \ using the \citealt{Saumon:2008im}
``hybrid'' models).

The slope of the RV curve and the projected separation of the companion 
provide direct information about the companion mass, enabling us to
readily test whether this brown dwarf hypothesis for 12 Psc B is compatible with the measured 
radial acceleration.
Following \citet{Torres:1999gc}, the mass of a companion accelerating its host star can be derived
by taking the time derivative of the radial velocity equation of a Keplerian orbit.
The companion mass ($M_B$) is related to the system distance ($d$),
the projected separation of the companion ($\rho$), the instantaneous slope of the radial  acceleration
($dv_r/dt$), and the orbital elements (eccentricity $e$, argument of periastron $\omega$, 
inclination $i$, and Keplerian angles related to the true orbital phase---the 
true anomaly $f$ and eccentric anomaly $E$) as follows:

\begin{eqnarray}{\label{eqn:dvdt}}
\frac{M_B}{\Msun} = 5.341 \times 10^{-6} \Big(\frac{d}{\mathrm{pc}}\Big)^2 \Big(\frac{\rho}{''}\Big)^2 \bigg| \Big(\frac{dv_r/dt}{\mathrm{m~s^{-1}~yr^{-1}}}\Big) \times   \nonumber \\
   (1-e)(1+\cos{E}) \times \bigg[ (1-e\cos E) \sin{(f + \omega)}  \nonumber \\
   \big(1 - \sin^2{(f + \omega)} \sin^2{i}\big)  (1 + \cos{f}) \sin i \bigg]^{-1}  \bigg| .
\end{eqnarray}

% Figure 7

\begin{figure*}
  \vskip -0.2 in
  \hskip -0.5 in
  \resizebox{8.5in}{!}{\includegraphics{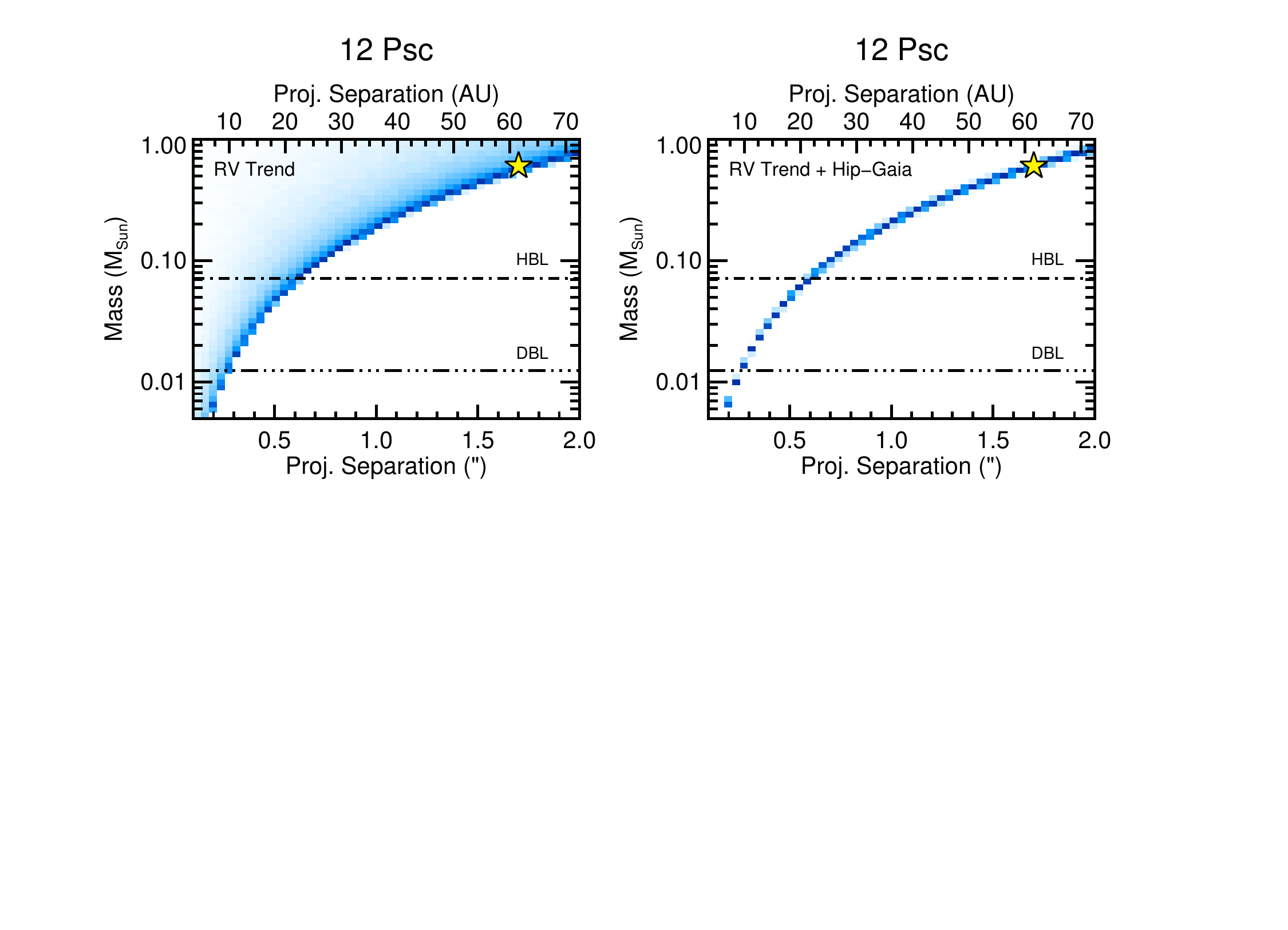}}
  \vskip -3 in
  \caption{Joint constraints on the mass and projected separation of a companion to 12 Psc based only on the measured
  RV trend (left) and using both the RV trend and the astrometric acceleration (right).
   The hydrogen burning limit (HBL; $\approx$75~\Mjup) and deuterium-burning limit ($\approx$13~\Mjup) are labeled.
   Based on the strength of the radial acceleration, the companion would be a brown dwarf or giant planet at close separations
   of $\lesssim$20~AU; more massive stellar or white dwarf companions are required on wider orbits.
   When the HGCA acceleration is included, the range of possible masses and separations is more limited.
   The measured separation and dynamical mass of 12 Psc B (from Section~\ref{sec:12psc_orbitfit}) is shown with the star.
   \label{fig:12psc_minmass} } 
\end{figure*}

% Figure 8

\begin{figure*}
  \vskip -0.2 in
  \hskip -0.5 in
  \resizebox{8.5in}{!}{\includegraphics{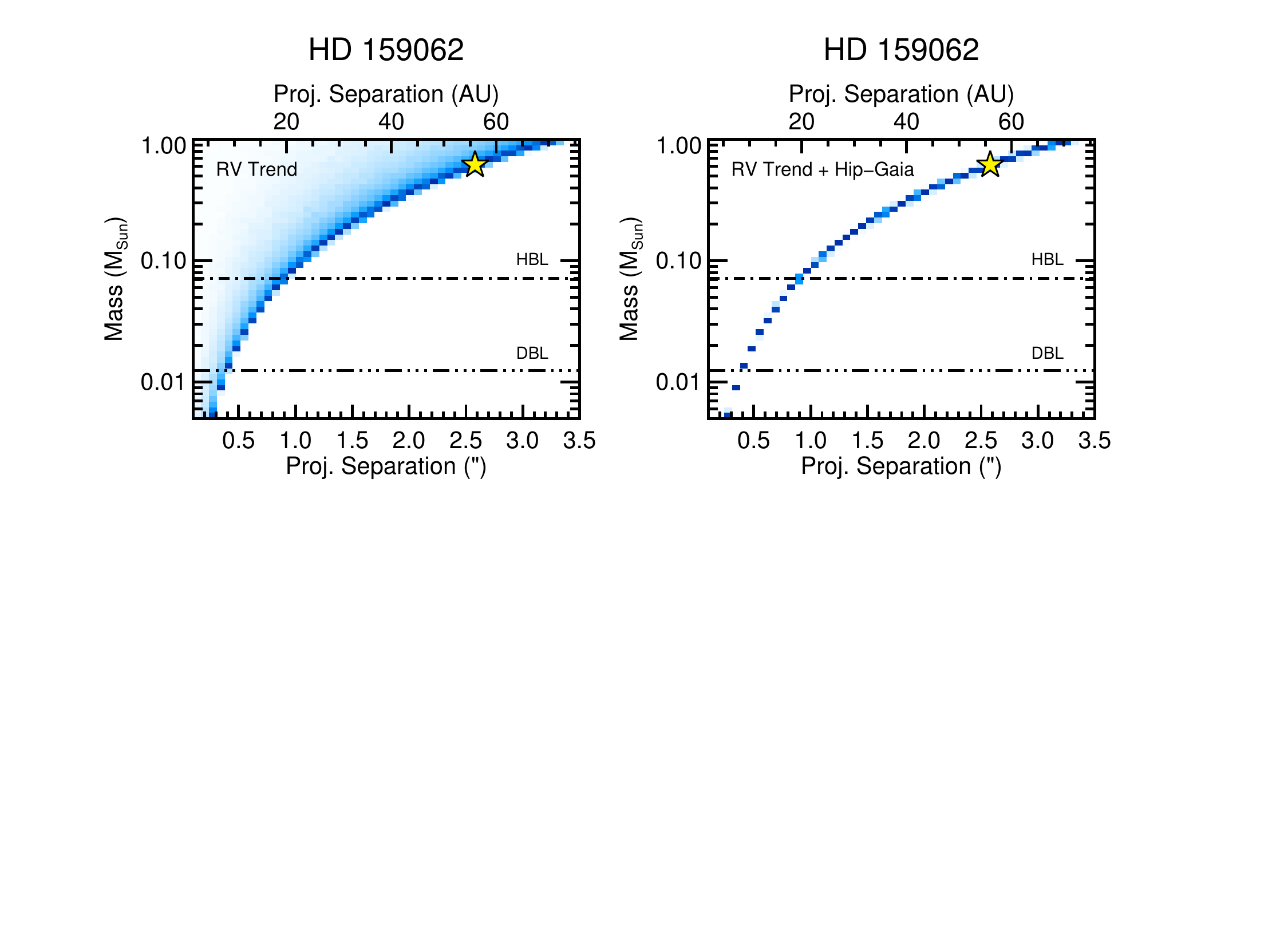}}
  \vskip -3 in
  \caption{Same as Figure~\ref{fig:12psc_minmass} but for HD 159062.  The strength of the radial and astrometric accelerations
  together imply that the companion has a mass above the stellar limit at separations beyond about 20 AU.
  The measured separation and dynamical mass of HD 159062 B are in good agreement with this prediction.  \label{fig:hd159062_minmass} } 
\end{figure*}

Using only the measured RV slope, 
the companion mass distribution can be constrained  
with reasonable assumptions about the distributions of
(\emph{a priori} unknown) orbital elements and orbital phase angles projected on the plane of the sky.
Here we adopt a uniform distribution for the argument of periastron from 0--2$\pi$,
uniform eccentricities between 0--1, and an isotropic distribution of inclination angles projected on the sky
(equivalent to a uniform distribution in $\cos i$).  To calculate the orbital phase,  
a mean anomaly is randomly drawn from 0--2$\pi$,
an eccentric anomaly is iteratively solved for using the Newton-Raphson method,
and a true anomaly is computed using 
$\tan(f/2) = \sqrt( (1+e)/(1-e)) \tan(E/2)$.  Repeating this process with Monte Carlo 
draws over a range of projected separations results in a joint probability distribution
between the companion mass and separation based on the measured radial acceleration
(and conditioned on our assumptions about the distribution of orbital elements).

Results for 12 Psc are shown in Figure~\ref{fig:12psc_minmass}.  The 
measured slope of $dv_r$/$d t$ = 10.60 $\pm$ 0.13 m s$^{-1}$ yr$^{-1}$  
implies that if the companion causing the acceleration 
was a brown dwarf or giant planet, it must be located at $\lesssim$0$\farcs$6 ($\lesssim$20~AU).
Assuming the RV trend originates entirely from the imaged companion 
we identify at 1$\farcs$6, 
the minimum mass of 12 Psc B is 0.49~\Msun.
This immediately rules out the brown dwarf scenario.
It also rules out any possibility that the companion could be a low mass star
because the faintest absolute magnitude this mass threshold corresponds to is $M_{K_S}$ $\approx$ 5.7~mag
following the empirical calibrations from \citet{Mann:2019ey}.
This is about 10 magnitudes brighter than the observed absolute magnitude, implying 
that 12 Psc B must be a white dwarf.

% Figure 9

\begin{figure*}
  \vskip -0 in
  \hskip -0.1 in
\epsscale{1.05}
  \plottwo{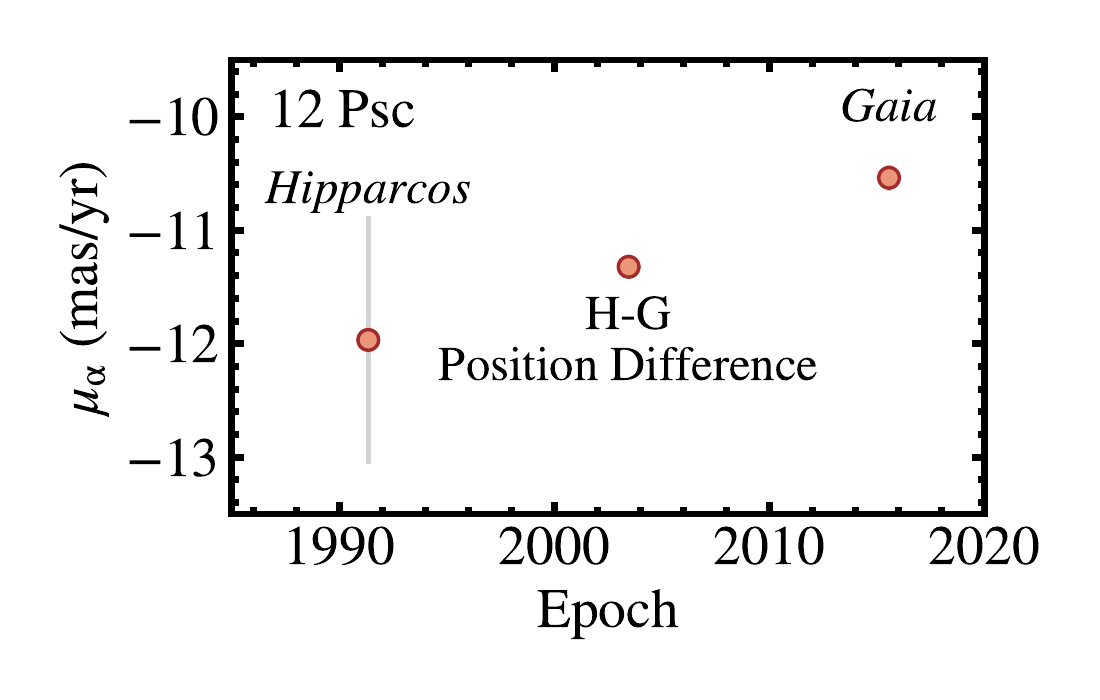}{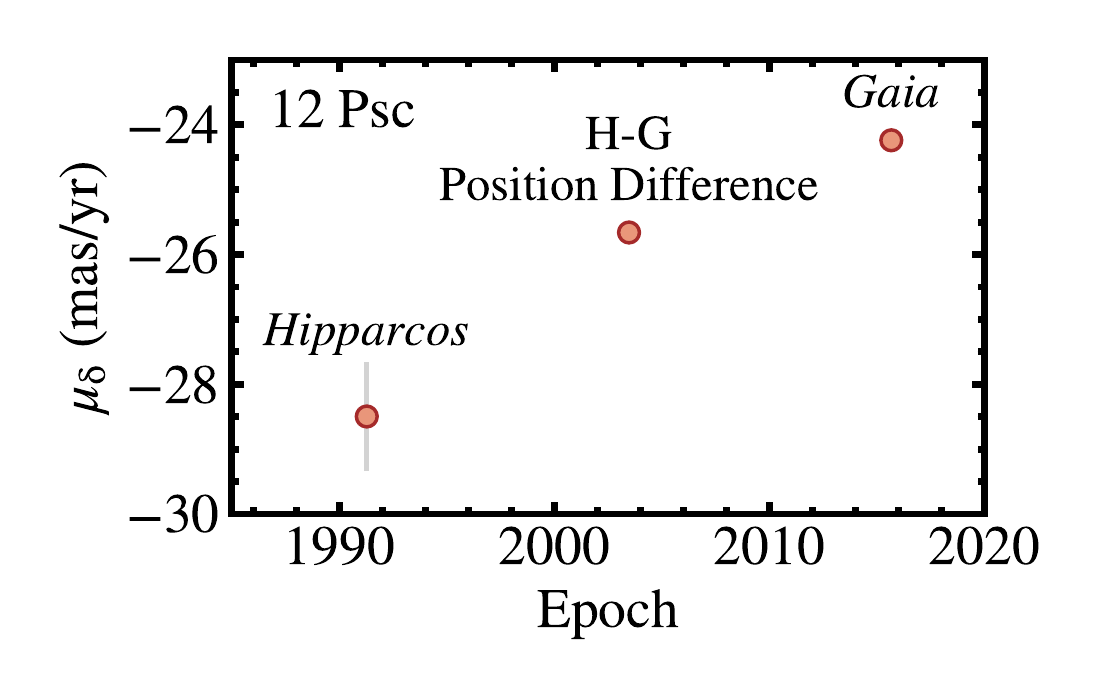}
  \hskip -0. in
\epsscale{1.05}
  \plottwo{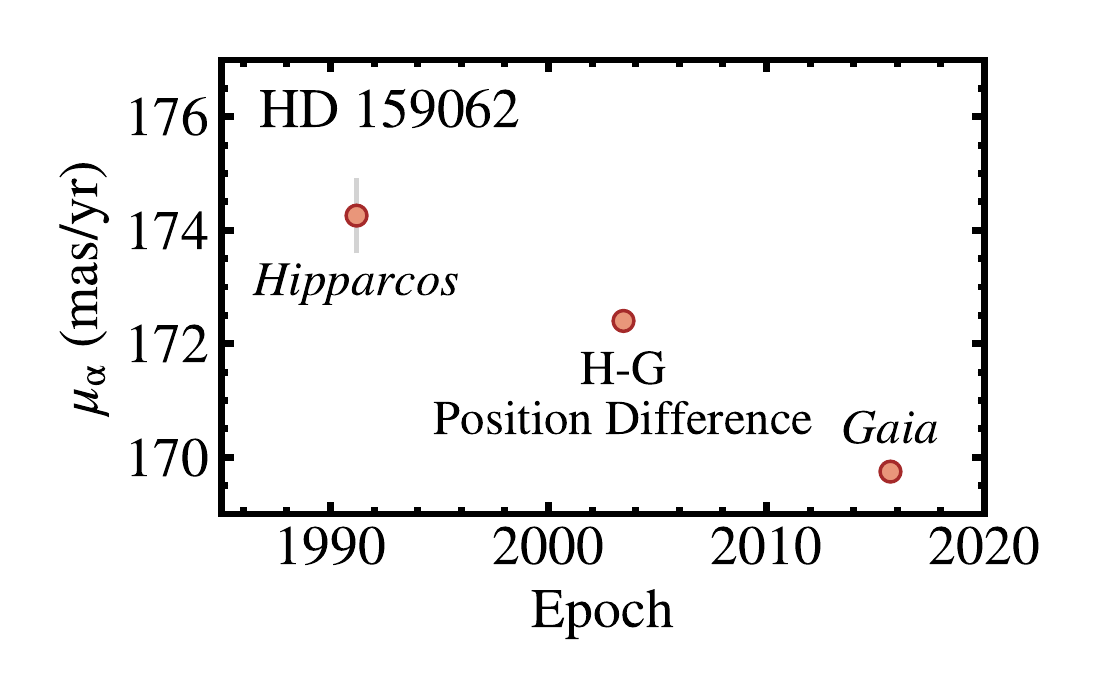}{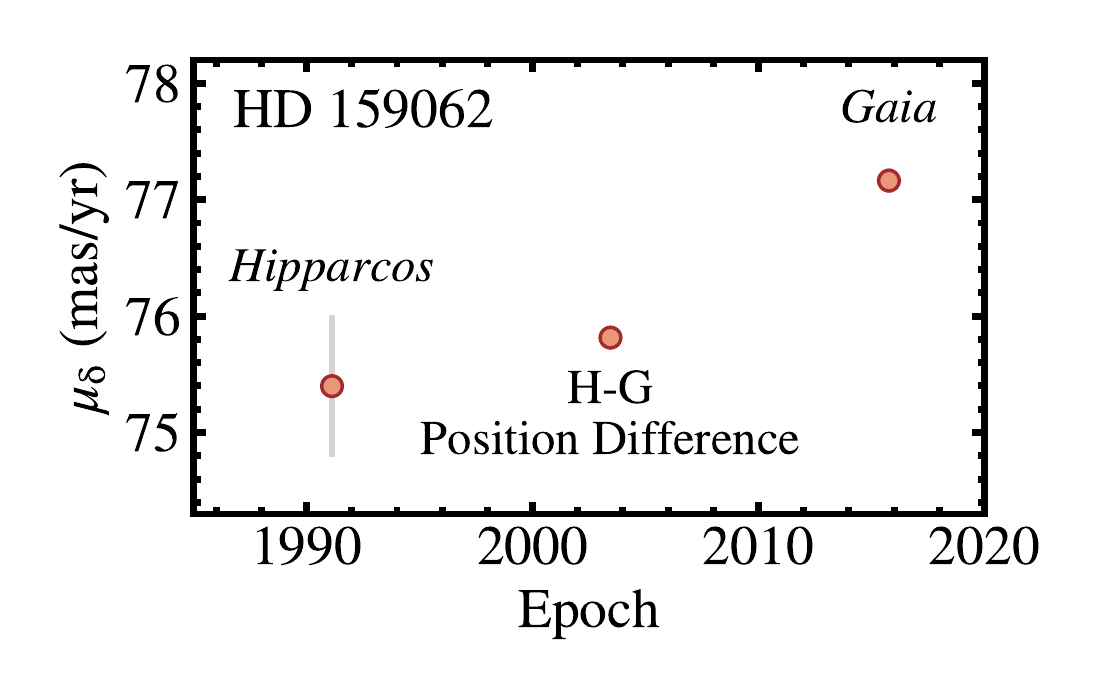}
  \vskip -0.1 in
  \caption{Proper motion measurements from HGCA in R.A. and Dec. for 
  12 Psc (top) and HD 159062 (bottom).  Three measurements are available: the ``instantaneous'' proper
  motion from \emph{Hipparcos}, a similar measurement from \emph{Gaia} DR2, and the scaled
  positional difference between the two missions (see \citealt{Brandt:2018dja} for details).  Both
  12 Psc and HD 159062 show clear changes in acceleration from 1991 to 2015; the slope of
  these proper motion measurements represents the astrometric acceleration.     \label{fig:pmradec} } 
\end{figure*}

\subsection{The Nature of HD 159062 B}

Following the same procedure as for 12 Psc, we show the joint mass-separation distribution
for HD 159062 in Figure~\ref{fig:hd159062_minmass} based on our measured RV trend
of $dv_r$/$d t$ = --14.1 $\pm$ 0.3 m s$^{-1}$ yr$^{-1}$.
This radial acceleration is only consistent with a brown dwarf or giant planet if the companion 
is located at a separation of $\lesssim$0$\farcs$9 ($\lesssim$20~AU); beyond this it must
be a low-mass star or a white dwarf.
At a separation of 2$\farcs$7, the minimum mass implied by the radial acceleration is 0.67~\Msun.
However, Equation~\ref{eqn:dvdt} assumes that the projected separation $\rho$ and slope $dv_r/dt$
are measured simultaneously.  The mean epoch of our RV measurements is 2011.5, which
is significantly earlier than our imaging observations.  The closest astrometric epoch to that date
from \citet{Hirsch:2019cp} is 2012.481, at which point HD 159062 B was located at 2$\farcs$594 $\pm$ 0$\farcs$014.
In Section~\ref{sec:cpm} we found that HD 159062 B is moving away from its host at a rate of
13.8~mas yr$^{-1}$.    If we correct for that average motion over the course of one year and 
assume a separation of 2$\farcs$580 $\pm$ 0$\farcs$014
in mid-2011, the minimum mass becomes 0.612 $\pm$ 0.015~\Msun, which takes into account 
uncertainties in the RV trend, distance, and projected separation.
\citet{Hirsch:2019cp} and Brandt et al. (submitted) 
analyzed observations of HD 159062 B in detail and unambiguously demonstrated that
this companion is a white dwarf.
Brandt et al. (submitted) find a dynamical mass of 0.617$^{+0.013}_{-0.012}$~\Msun \ for HD 159062 B,
which is in good agreement with our inferred minimum mass from the RV slope.

\input{wd_12psc_mk_orbit_table.tex}

\subsection{Hipparcos-Gaia Accelerations}{\label{sec:hgca}}

\citet{Brandt:2018dja} carried out a cross calibration between the \emph{Hipparcos} and \emph{Gaia}
astrometric datasets which resulted in the \emph{Hipparcos}-\emph{Gaia} Catalog of Accelerations (HGCA).
Linking these catalogs to a common reference frame (that of \emph{Gaia} DR2) provides a way to 
correct for local sky-dependent systematics present the \emph{Hipparcos}.  As a result, measurements of 
astrometric accelerations between these two missions (separated by $\approx$25 years) can be inferred
through changes in proper motion.  The HGCA has been an especially valuable tool to 
measure dynamical masses of long-period substellar companions by combining absolute accelerations with
relative astrometry and radial velocities
(\citealt{Brandt:2019ey}; \citealt{Dupuy:2019cy}; \citealt{Brandt:2019kp}; Franson et al., in prep.).

Both 12 Psc and HD 159062 have significant astrometric accelerations in HGCA (Figure~\ref{fig:pmradec}).  
Three proper motions in R.A. and Dec. are available in this catalog: the proper motion in \emph{Hipparcos}
with a mean epoch of 1991.25, the proper motion  in \emph{Gaia} with a mean epoch of 2015.5, and
the scaled positional difference between \emph{Hipparcos} and \emph{Gaia}.  The latter measurement is the most
precise as a result of the long baseline between the two missions.

We compute astrometric accelerations in R.A. ($d \mu_{\alpha}/dt$) and Dec. ($d \mu_{\delta}/dt$)  
using the proper motion from \emph{Gaia} and the inferred proper motion 
from the scaled \emph{Hipparcos}-\emph{Gaia} positional difference following \citet{Brandt:2019ey}: 
$d \mu_{\alpha}/dt$ = 2 $\Delta \mu_{\alpha, \mathrm{Gaia-HG}}$ / $(t_{\alpha, \mathrm{Gaia}} - t_{\alpha, \mathrm{Hip}})$ and 
$d \mu_{\delta}/dt$ = 2 $\Delta \mu_{\delta, \mathrm{Gaia-HG}}$ / $(t_{\delta, \mathrm{Gaia}} - t_{\delta, \mathrm{Hip}})$.
Here $t_{\alpha, \mathrm{Gaia}}$ and $t_{\delta, \mathrm{Gaia}}$ are the \emph{Gaia} astrometric epochs
corresponding to the proper motion measurements in R.A. and Dec., and $t_{\alpha, \mathrm{Hip}}$ and $t_{\delta, \mathrm{Hip}}$
are the corresponding epochs for \emph{Hipparcos}.
The total  acceleration is then computed as $d \mu_{\alpha \delta}/dt = \sqrt{(d \mu_{\alpha}/dt)^2  + (d \mu_{\delta}/dt)^2}$.

\citet{Brandt:2019ey} presented a simple relationship between the 
mass of a companion ($M_B$), its instantaneous projected separation ($\rho$), 
and both the radial ($dv_r$/$dt$) and astrometric ($d \mu_{\alpha \delta}/dt$) accelerations induced on its host star.
In convenient units this can be expressed as

\begin{eqnarray}{\label{eqn:hgca_mass_mjup}}
\frac{M_B}{\Mjup} = 5.599 \times 10^{-3} \Big(\frac{d}{\mathrm{pc}}\Big)^2 \Big(\frac{\rho}{''}\Big)^2 
\Big[ \Big(\frac{d\mu_{\alpha \delta}/dt}{\mathrm{m~s^{-1}~yr^{-1}}}\Big)^2 +    \nonumber \\
\Big(\frac{dv_r/dt}{\mathrm{m~s^{-1}~yr^{-1}}} \Big)^2 \Big]^{3/2} 
\Big(\frac{d\mu_{\alpha \delta}/dt}{\mathrm{m~s^{-1}~yr^{-1}}}\Big)^{-2}.
\end{eqnarray}

% Figure 10

\begin{figure*}
  \vskip -1.8 in
  \hskip 0.5 in
  \resizebox{6.in}{!}{\includegraphics{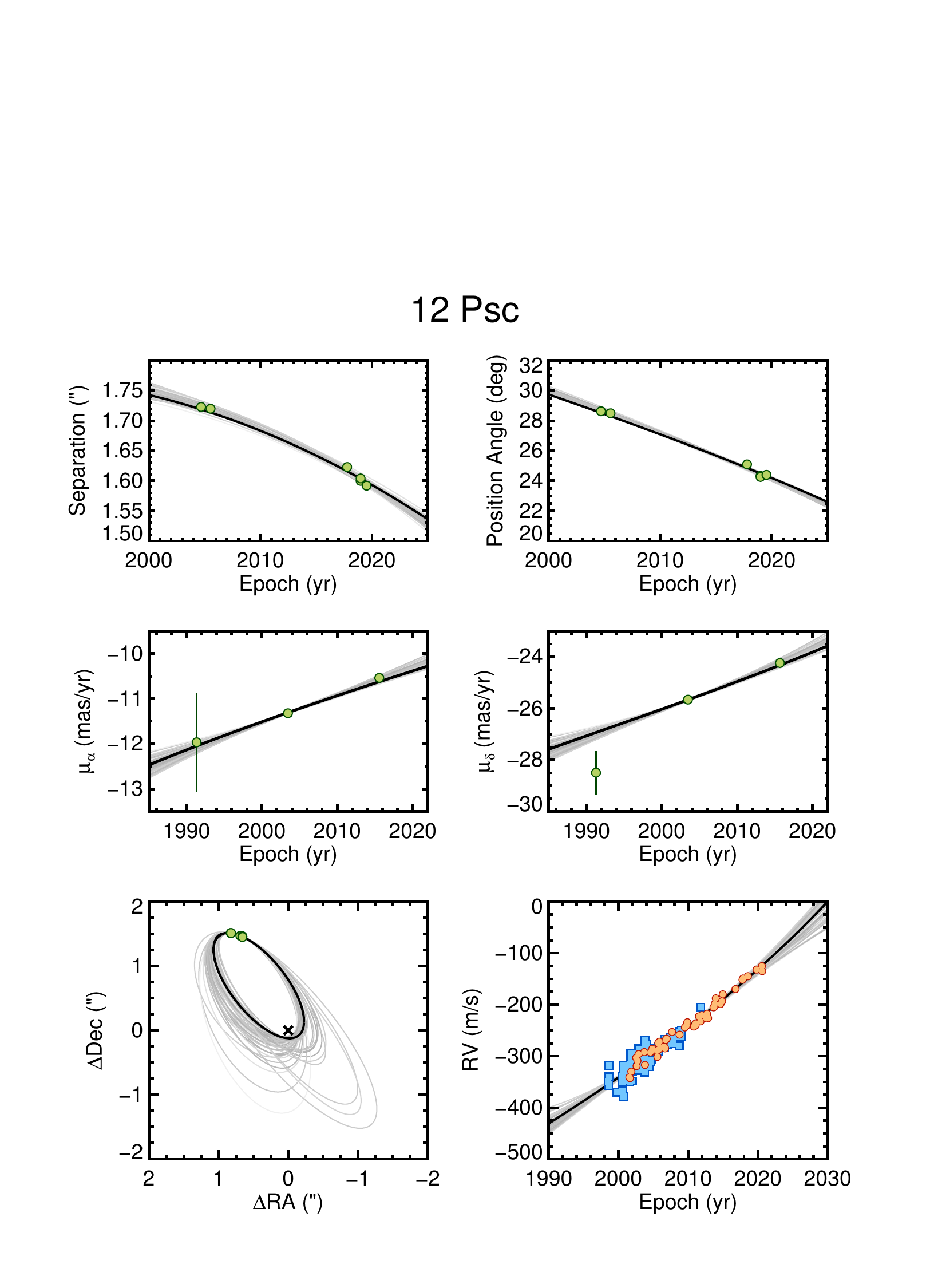}}
  \vskip -0.4 in
  \caption{Keplerian orbit fit to the relative astrometry of 12 Psc B (top and lower left panels), 
  the astrometric acceleration of the host star from the HGCA (middle panels), 
  and RVs of 12 Psc (lower right panel). 
  50 random orbits drawn from the MCMC chains are shown in gray and are color coded based on their
  $\chi^2$ values; darker gray indicates a lower $\chi^2$ value and a better fit.  The best fit
  orbit is shown in black. 
  In the lower right panel, blue squares are Lick RVs and orange circles are from the Tull Spectrograph.
  \label{fig:12psc_orbitfits} } 
\end{figure*}

% Figure 11

\begin{figure*}
  \vskip -0. in
  \hskip -0.3 in
  \resizebox{7.5in}{!}{\includegraphics{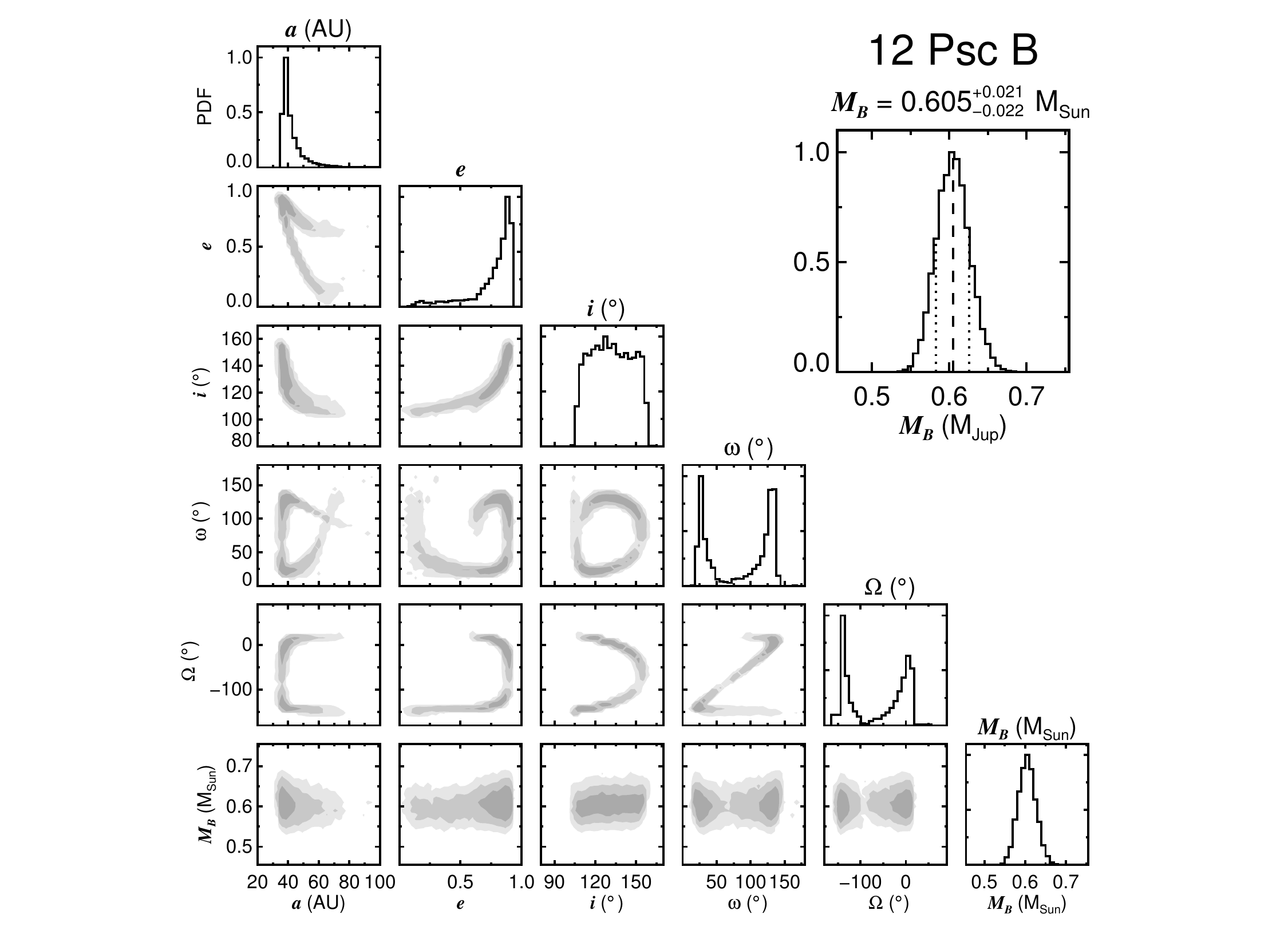}}
  \vskip -0 in
  \caption{Corner plot showing joint posterior maps of various parameters and their marginalized probability density functions.
  The observations of 12 Psc B and its host star favor a high eccentricity orbit and a semi-major axis of $\approx$40~AU.  
  The dynamical mass of the white dwarf companion 12 Psc B 
  is 0.605$^{+0.021}_{-0.022}$ \Msun.  \label{fig:12psc_corner} } 
\end{figure*}

\noindent The numerical coefficient becomes 5.342 $\times$ 10$^{-6}$ when $M_B$ is in units of \Msun.
Equation~\ref{eqn:hgca_mass_mjup} is valid when all three measurements 
($\rho$, $dv_r$/$dt$, and $d \mu_{\alpha \delta}/dt$) are obtained simultaneously.
This is rarely the case in practice, but this relation offers a convenient approximation of the dynamical mass
as long as the orbit has not evolved substantially, as is the case for long-period companions.

The HGCA kinematics for 12 Psc are listed in Table~\ref{tab:12pschostprop} and displayed in Figure~\ref{fig:pmradec}.
12 Psc shows a significant change in proper motion between the 
\emph{Hipparcos}-\emph{Gaia} scaled positional difference and \emph{Gaia} measurements: 
$\Delta \mu_{\alpha, \mathrm{Gaia-HG}}$ = 0.78 $\pm$ 0.12 mas yr$^{-1}$ and
$\Delta \mu_{\delta, \mathrm{Gaia-HG}}$ = 1.42 $\pm$ 0.09 mas yr$^{-1}$.
This translates into an astrometric acceleration of $d \mu_{\alpha \delta}/dt$ = 22.9 $\pm$ 1.4 m s$^{-1}$ yr$^{-1}$---about 
twice as large as its radial acceleration.  
The inferred constraints on the companion mass and separation  resulting from these radial and astrometric
accelerations are shown in Figure~\ref{fig:12psc_minmass}.
The projected separation of 12 Psc ranges from 1$\farcs$723 in 2004 to
1$\farcs$592 in 2019.  Using a projected separation of $\rho$=1$\farcs$7, which is closer to the mid-points of the radial
and astrometric accelerations, Equation~\ref{eqn:hgca_mass_mjup} implies a mass of about 
0.622 $\pm$ 0.018~\Msun.  This is typical of white dwarf masses (e.g., \citealt{Kepler:2007jz}).

HD 159062 also shows substantial changes in proper motion in the HGCA (Figure~\ref{fig:pmradec}):
$\Delta \mu_{\alpha, \mathrm{Gaia-HG}}$ = --2.66 $\pm$ 0.12 mas yr$^{-1}$ and
$\Delta \mu_{\delta, \mathrm{Gaia-HG}}$ = 1.35 $\pm$ 0.11 mas yr$^{-1}$.
This translates into an astrometric acceleration of $d \mu_{\alpha \delta}/dt$ = 25.0 $\pm$ 1.0 m s$^{-1}$ yr$^{-1}$.
Figure~\ref{fig:hd159062_minmass} shows constraints combining this 
with the RV trend.  At a separation of 2$\farcs$580 $\pm$ 0.014 mas,
the implied mass HD 159062 B is 0.632 $\pm$ 0.014~\Msun---in good agreement
with the dynamical mass.
Below we carry out a full orbit fit of 12 Psc B and HD 159062 B using relative astrometry,
radial velocities, and absolute astrometry from the HGCA.

\subsection{Orbit and Dynamical Mass of 12 Psc B}{\label{sec:12psc_orbitfit}}

The orbit and dynamical mass of 12 Psc B are determined using the efficient orbit fitting package
\texttt{orvara} (Brandt et al., submitted), which jointly fits Keplerian orbits to radial velocities,
relative astrometry of resolved companions, and absolute astrometry from HGCA.
The code relies on a Bayesian framework with the \texttt{emcee} affine-invariant implementation of 
Markov Chain Monte Carlo (\citealt{ForemanMackey:2013io})
to sample posterior distributions of orbital elements, physical parameters
of the host and companion, and nuisance parameters like stellar parallax, instrument-dependent 
RV offsets, and RV jitter.
We use 100 walkers with 10$^5$ steps for our orbit fit of the 12 Psc system.

Our priors are chosen to ensure the observations
drive the resulting posteriors.  We adopt log-flat priors for the semi-major axis, 
companion mass, and RV jitter; a $\sin i$ distribution for inclination; and linearly uniform priors 
for all other fitted parameters ($\sqrt{e} \sin \omega$, $\sqrt{e} \cos \omega$, longitude of ascending node, 
and longitude at a reference epoch).
A Gaussian prior with a mean of 1.1 and a standard deviation of 0.2~\Msun \ is chosen for the
primary mass based on independent estimates from the literature 
(e.g. 1.11 $\pm$ 0.05~\Msun, \citet{Soto:2018bl};
1.079 $\pm$ 0.012~\Msun, \citealt{Tsantaki:2013dc};
1.12 $\pm$ 0.08~\Msun, \citealt{Mints:2017di}).

Results of the orbit fit for 12 Psc B are shown in 
Figures~\ref{fig:12psc_orbitfits} and \ref{fig:12psc_corner}\footnote{Note that the parameter 
values for orbit fits quoted in this study represent the median of the 
parameter posterior distributions and the 68.3\% credible interval, 
although we also list the best-fit values and the maximum 
\emph{a posteriori} probabilities in Table~\ref{tab:12pscb_orbit}.}.
The orbit of 12 Psc B has a high eccentricity of 0.84$\pm$0.08, an orbital period of 193$^{+25}_{-38}$~yr,
and a semi-major axis of 39.5$^{+2.8}_{-3.5}$~AU.
The dynamical mass of 12 Psc B is 0.605$^{+0.021}_{-0.022}$ \Msun, which is similar to our mass estimate
in Section~\ref{sec:hgca} using simplifying assumptions.
A summary of prior and posterior distributions for relevant fitted parameters can be found in Table~\ref{tab:12pscb_orbit}.

\input{wd_hd159062_mk_orbit_table.tex}

\subsection{Orbit and Dynamical Mass of HD 159062 B}

We fit a Keplerian orbit using \texttt{orvara} jointly to our HRS RVs, our new NIRC2 astrometry, and the HGCA acceleration for HD 159062 together with 
HIRES RVs and relative astrometry from \citet{Hirsch:2019cp}.
The same priors we used for 12 Psc B in Section~\ref{sec:12psc_orbitfit} are adopted for HD 159062 B except for 
the host star mass.  For this we use a Gaussian prior with a mean of 0.8 and a standard deviation of 0.2~\Msun, which captures
the typical range of mass estimates for HD 159062 from the literature 
(e.g., 0.76 $\pm$ 0.03~\Msun, \citealt{Hirsch:2019cp};
0.78 $\pm$ 0.03~\Msun, \citealt{Mints:2017di};
0..87~\Msun, \citealt{Luck:2017jd}).

\begin{figure*}
  \vskip -1.8 in
  \hskip 0.4 in
  \resizebox{6.in}{!}{\includegraphics{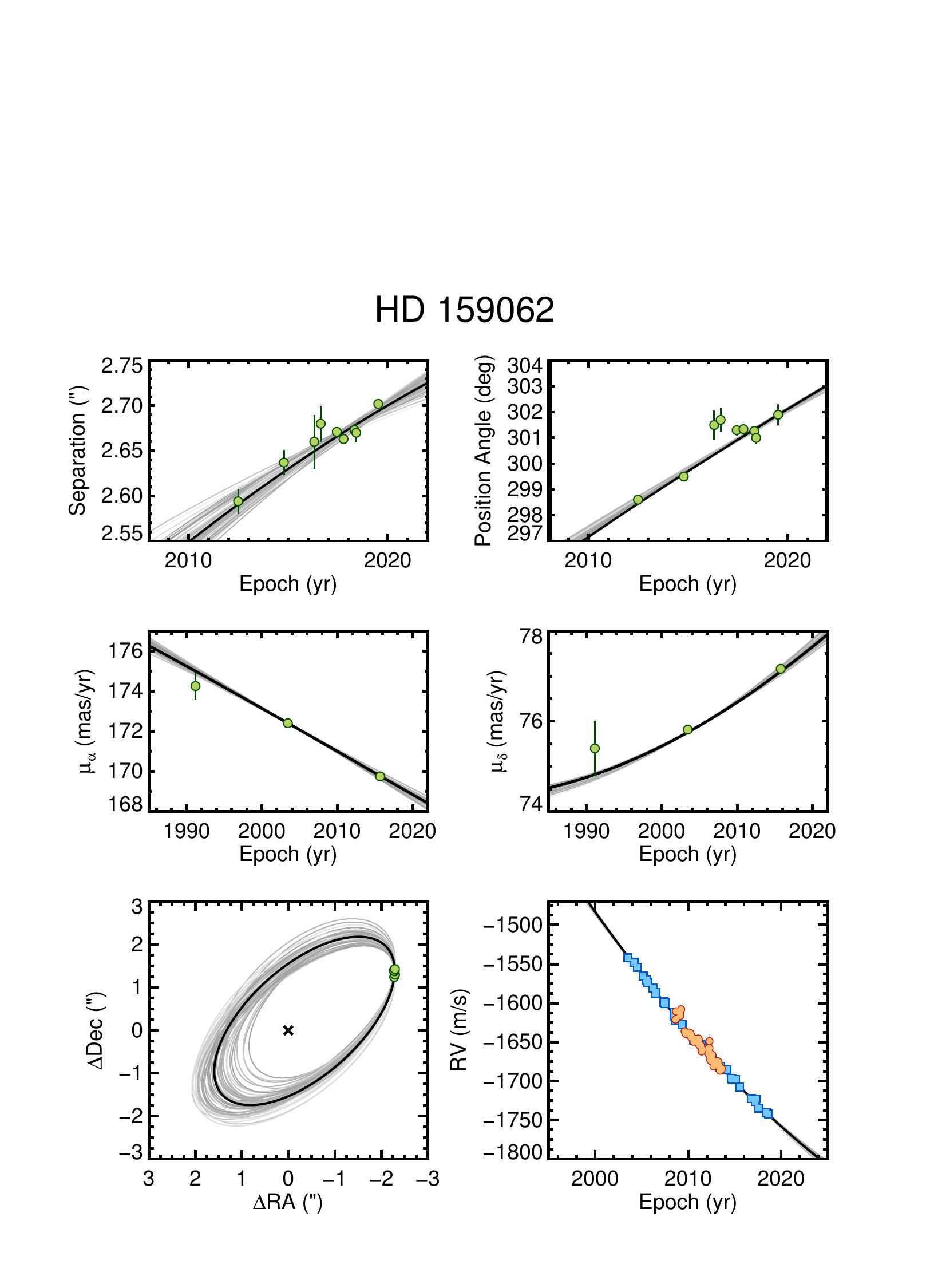}}
  \vskip -0.4 in
  \caption{Keplerian orbit fit for HD 159062 B from relative astrometry (top), astrometric acceleration from HGCA (middle panels), 
  and radial velocities (bottom right).  See Figure~\ref{fig:12psc_orbitfits} for details.  In the lower right panel, our RVs from HRS are shown
  as orange circles while HIRES RVs from \citet{Hirsch:2019cp} are shown as blue squares. \label{fig:hd159062_orbitfits} } 
\end{figure*}

\begin{figure*}
  \vskip -0. in
  \hskip -0.3 in
  \resizebox{7.5in}{!}{\includegraphics{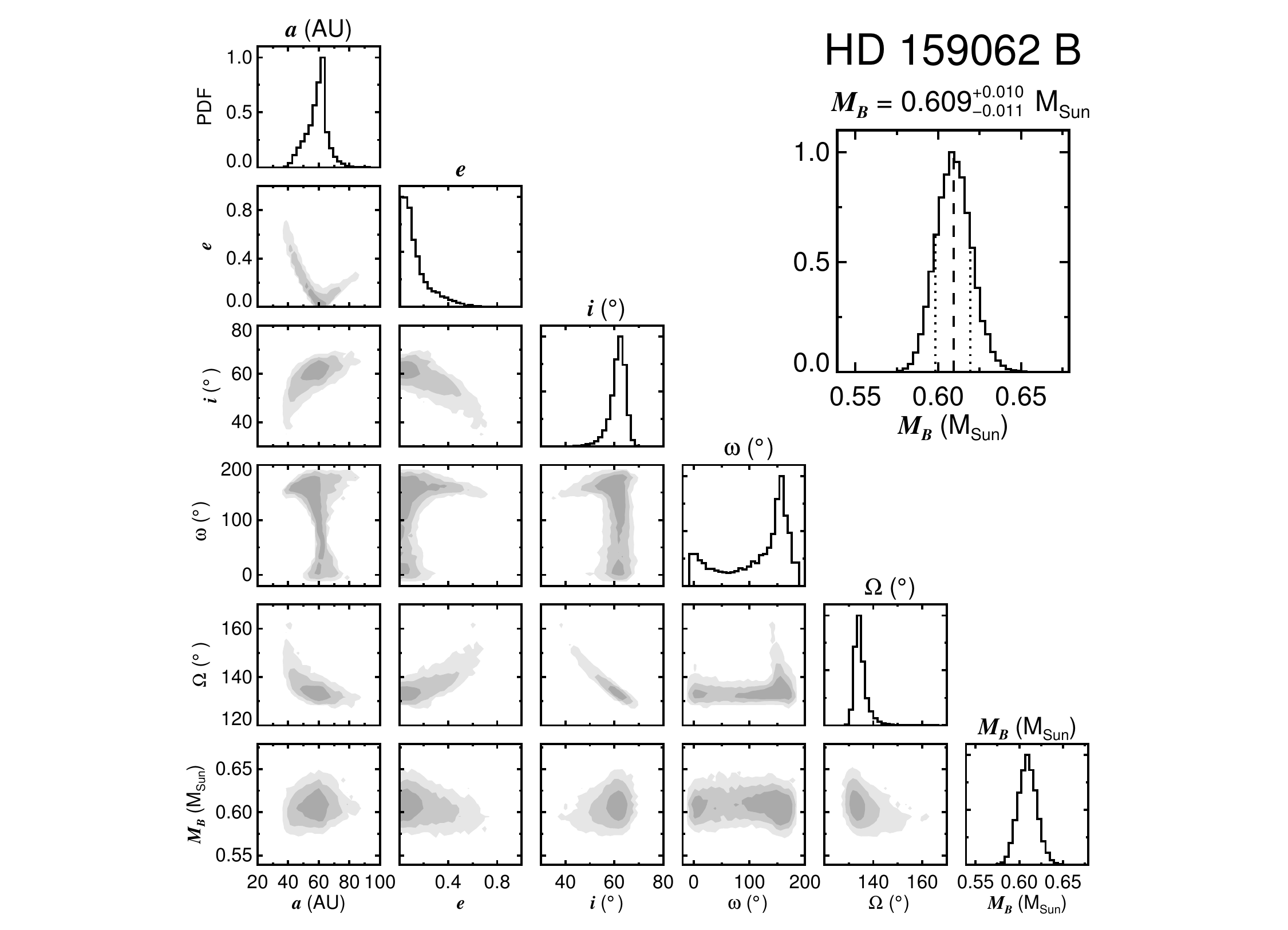}}
  \vskip -0 in
  \caption{Same as Figure~\ref{fig:12psc_corner} but for HD 159062 B.  The dynamical mass of the white dwarf companion HD 159062 B 
  is 0.609$^{+0.010}_{-0.011}$ \Msun.  \label{fig:hd159062_corner} } 
\end{figure*}

Results of the orbit fit are shown in Figures~\ref{fig:hd159062_orbitfits} and \ref{fig:hd159062_corner}, and a summary of the posterior distributions 
is listed in Table~\ref{tab:hd159062b_orbit}.
The dynamical mass of HD 159062 B is 0.609$^{+0.010}_{-0.011}$ \Msun, 
which happens to be very similar to the mass we found for 12 Psc B.  
HD 159062 B orbits with a semi-major axis of 60$^{+5}_{-7}$ AU, a period of
390 $\pm$ 70 yr, and a low eccentricity which
peaks at $e$=0.0 and is below $e$=0.42 with 95\% confidence.
These are consistent with but more precise than the values found by \citealt{Hirsch:2019cp} and Brandt et al. (submitted).

Our dynamical mass of 0.609$^{+0.010}_{-0.011}$ \Msun \ is 
somewhat lower than
the value of 0.65$^{+0.12}_{-0.04}$~\Msun \ from \citet{Hirsch:2019cp}
and 0.617$^{+0.013}_{-0.012}$~\Msun \ from Brandt et al. (submitted).
Because the mass of white dwarf remnants scales strongly with progenitor mass, a lower final mass implies
a substantially lower initial mass (and longer main-sequence lifetime) 
compared to the 2.4~\Msun \ progenitor mass found by \citealt{Hirsch:2019cp}.
The implications of a lower progenitor mass are discussed in more detail below.

% Figure 12

% Figure 13

\section{Discussion and Conclusions}{\label{sec:discussion}}

Stars with masses $\lesssim$8~\Msun \ 
evolve to become white dwarfs on timescales of a few tens of Myr 
for high-mass stars near the threshold for core-collapse supernovae (e.g., \citealt{Ekstrom:2012ke}; \citealt{Burrows:2013hpa})
to $\sim$10$^{4}$~Gyr for the lowest-mass stars near the hydrogen-burning limit (\citealt{Laughlin:1997ht}).
Given the 13.8~Gyr age of the universe, the lowest-mass stars that can have evolved in isolation to become white dwarfs 
have masses of $\approx$0.9--1~\Msun.
White dwarf masses generally increase with progenitor mass, and the 
corresponding minimum white dwarf mass that can result from 
isolated evolution of such a star at solar metallicity is $\gtrsim$0.56~\Msun \ (\citealt{Cummings:2018cz}).
Most white dwarfs should therefore have masses above this value, and indeed the peak
of the white dwarf mass function in the solar neighborhood is $\approx$0.6~\Msun \ (e.g., \citealt{Liebert:2005go}; \citealt{Kepler:2007jz}).
The majority of these have hydrogen atmospheres with DA classifications (\citealt{Kepler:2007jz}).

The initial-to-final mass relation connects a progenitor star's mass to the final mass of the resulting white dwarf remnant. 
These relations are sensitive to metallicity and the physics of AGB evolution, 
including shell burning, dedge-up events, and mass loss (e.g., \citealt{Marigo:2007cw}).
\citet{Cummings:2018cz} provide a semi-empirical calibration of the white dwarf initial-to-final mass relation 
spanning initial masses of $\approx$0.9--7~\Msun, or final masses between $\approx$0.5--1.2~\Msun.
Using this relation, our dynamical mass measurements for the white dwarfs 12 Psc B and HD 159062 B imply nearly identical initial progenitor masses
of 1.5 $\pm$ 0.6~\Msun \ for both companions.
The large uncertainties reflect the significant scatter in the empirically calibrated initial-to-final mass relation, which is 
sparsely populated for initial masses $\lesssim$2.5~\Msun.
Theoretical stellar evolutionary models also exhibit substantial dispersion in the initial-to-final mass relation, with initial
masses predicted to be between $\approx$1.3--2.2~\Msun \ at solar metallicity for the final masses we measure for 12 Psc B
and HD 159062 B (e.g., \citealt{Marigo:2007cw}; \citealt{Choi:2016kf}).

The 12 Psc system was therefore initially a $\sim$1.5~\Msun \ + 1.1~\Msun \ binary and 
the HD 159062 system was initially a $\sim$1.5~\Msun \ + 0.8~\Msun \ binary.
The more massive components then underwent standard evolution through the giant and AGB phases.
At this point, after about 2.9 Gyr of evolution, their radii expanded to $\sim$450 $R_{\odot}$ ($\sim$2.1~AU) before
shedding $\approx$60\% of their initial mass to become white dwarfs 
(\citealt{Paxton:2010jf}; \citealt{Choi:2016kf}; \citealt{Dotter:2016fa}; \citet{Cummings:2018cz}).
Adopting an age of 5.3 $\pm$ 1.1 Gyr for 12 Psc from \citet{Soto:2018bl}, the most likely companion progenitor mass of $\sim$1.5~\Msun \ 
implies that the cooling age of 12 Psc B is $\sim$2--3~Gyr.  A higher (lower) progenitor mass would result in a longer (shorter) cooling age.
The age of HD 159062 is somewhat more uncertain, but for a system age of $\sim$9--13~Gyr,
a main-sequence lifetime of $\approx$3~Gyr for the progenitor of
HD 159062 B implies a cooling time of $\sim$6--10 Gyr for the white dwarf.
This is consistent with the cooling age of 8.2$^{+0.3}_{-0.5}$~Gyr derived by \citet{Hirsch:2019cp}.

\begin{deluxetable*}{lcccccccc}
\renewcommand\arraystretch{0.9}
\tabletypesize{\small}
\setlength{ \tabcolsep } {.1cm} 
\tablewidth{0pt}
\tablecolumns{9}
\tablecaption{Resolved ``Sirius-Like'' White Dwarf Companions with Dynamical Mass Measurements\label{tab:wdcomps}}
\tablehead{
                                  & \colhead{Dynamical Mass}  & \colhead{White Dwarf} & \colhead{Host Star} &  \colhead{Proj. Sep.\tablenotemark{a}}  &  \colhead{$a$}  & \colhead{System Age}  &  \colhead{WD Cooling Age}  &          \\
       \colhead{Name} & \colhead{(\Msun)}                 & \colhead{SpT}             & \colhead{SpT}          &  \colhead{($''$)}          &  \colhead{(AU)}           & \colhead{(Gyr)}             & \colhead{(Gyr)}                     & \colhead{Ref.} 
        }   
\startdata
40 Eri B  & 0.573 $\pm$ 0.018  & DA2.9 &   M4.5+K0  & 8.3 &  35  & $\approx$1.8 &  $\approx$0.122  &  1,  2 , 3 \\
Procyon B & 0.592 $\pm$ 0.006 & DQZ & F5 IV--V & 3.8 & 15 & $\sim$2.7 & 1.37 $\pm$ 0.04 &  4 \\
Gl 86 B &  0.596 $\pm$ 0.010 & DQ6 &  K0 V & 2.4 & 22  & $\sim$2.5 & 1.25 $\pm$ 0.05   & 5, 6  \\
12 Psc B  & 0.605$^{+0.021}_{-0.022}$ & $\cdots$  & G1 V & 1.6 & 40 & 5.3 $\pm$ 1.1 & $\sim$2--3  &  7 \\
HD 159062 B  & 0.609$^{+0.010}_{-0.011}$ & $\cdots$ &  G9 V  &  2.7  & 60 & $\sim$9--13 &  8$^{+3}_{-5}$  &  7, 8  \\
Stein 2051 B\tablenotemark{b}  & 0.675 $\pm$ 0.051 &  DC & M4 & 10.1 & 56 & 1.9--3.6 & 1.9 $\pm$ 0.4 & 9  \\
Sirius B & 1.018  $\pm$ 0.011 & DA2 & A1 V & 10.7  & 20 & 0.288 $\pm$ 0.010 & $\approx$0.126 & 10 \\
\enddata
\tablecomments{Entries in this table are limited to white dwarf companions with precise dynamical mass constraints ($\sigma_M$/$M$$<$10\%).
}
\tablerefs{
(1) \citet{Gianninas:2011bp}; 
(2) \citet{Mason:2017ey};
(3) \citet{Bond:2017cx};
(4) \citet{Bond:2015doa};
(5) \citet{Farihi:2013ht};
(6) \citet{Brandt:2019ey}; 
(7) This work;
(8) \citet{Hirsch:2019cp};
(9) \citealt{Sahu:2017gs};
(10)  \citet{Bond:2017hca}.
}
\tablenotetext{a}{Most recently reported projected separation.}
\tablenotetext{b}{The mass of Stein 2051 B was measured via gravitational deflection.}
\end{deluxetable*}

Despite both binaries having broadly similar physical characteristics, it is interesting to note the differences in the orbits of these companions.
12 Psc B has a high eccentricity of $e$=0.84$\pm$0.08 and a semi-major axis of 40$^{+2}_{-4}$~AU,
which takes it to a periastron distance of 6.4 AU (with a 68\% credible interval spanning 2.7--10.2~AU).
On the other hand, HD 159062 B has a low eccentricity most consistent with a circular orbit ($e$$<$0.42~ at 95\% confidence),
a semi-major axis of 60$^{+5}_{-7}$~AU, and a periastron distance of 56$^{+8}_{-7}$ AU.
Given the orbital properties of 12 Psc B,
tidal interactions during the AGB phase 
should have been important for this system.  Without other mechanisms to increase eccentricities, these interactions
tend to dampen eccentricities and reduce orbital periods (e.g., \citealt{Saladino:2019bv}).
\citet{BonacicMarinovic:2008jza} highlight Sirius as an example of a binary which began as a $\sim$2.1+5.5~\Msun \ pair
which should have circularized, but Sirius B now orbits with an eccentricity of $e$=0.59 and an orbital period of 50~yr.
Assuming a mass ratio of $q$=$M_1$/$M_2$$\sim$1.4 for the unevolved 12 Psc system, 
the Roche lobe for the 12~Psc B progenitor would have been $\approx$2.6~AU at periastron 
following the approximation for the Roche lobe effective radius from \citet{Eggleton:1983aa}.
This is comparable to the size of 12 Psc B during the AGB phase ($\sim$2.1~AU).  
Mass transfer via Roche lobe overflow may therefore have occurred in this system,
and tidal interactions would have been important.
Like the Sirius system, the high eccentricity of 12 Psc B is therefore surprising and lends support to an eccentricity pumping mechanism, 
perhaps through enhanced mass loss at periastron which may counteract tidal circularization (\citealt{BonacicMarinovic:2008jza}).

Wide companions will evolve as if in isolation whereas short-period systems will evolve through one of several channels as detached,
semi-detached, or contact binaries.  12 Psc B and HD 159062 B occupy an intermediate regime at several tens of AU
where direct mass transfer via Roche lobe overflow may not have occurred, but wind accretion could have been important as a source
of chemical enrichment of the unevolved companion
as each progenitor underwent mass loss.
This is especially true for HD 159062, which shows
enhanced abundance of barium and other $s$-process elements---a signpost of 
prior accretion from an AGB companion (e.g., \citealt{Mcclure:1980aa}; \citealt{Escorza:2019cm}).
For example, \citet{Fuhrmann:2017aa} found a barium abundance of [Ba/Fe] = +0.40 $\pm$ 0.01 dex, and 
\citet{Reddy:2006hx} found 
[Y/Fe] = +0.37 dex, 
[Ce/Fe] = +0.10 dex, 
and [Nd/Fe] = 0.39 dex.
This led \citet{Fuhrmann:2017aa} to predict that HD 159062 harbors a white dwarf companion, which was later confirmed with
the discovery of HD 159062 B by \citet{Hirsch:2019cp}.

On the other hand, 12 Psc shows no signs of barium enrichment or significant enrichment from other $s$-process elements: 
\citet{DelgadoMena:2017cb} found abundances of 
[Ba/Fe] = --0.03 $\pm$ 0.02 dex, 
[Sr/Fe] = +0.12 $\pm$ 0.03 dex, 
[Y/Fe] = +0.09 $\pm$ 0.04 dex, 
[Zr/Fe] = +0.01 $\pm$ 0.08 dex, 
[Ce/Fe] = --0.07 $\pm$ 0.04 dex, 
and [Nd/Fe] = --0.09 $\pm$ 0.03 dex.
This lack of enrichment is surprising when compared to HD 159062: both have  
white dwarf companions with similar masses and presumably similar evolutionary pathways for their progenitors, but
12 Psc B is on a highly eccentric orbit which brings it much closer to its unevolved main-sequence companion 
at periastron ($\approx$6 AU for 12 Psc B versus $\approx$56~AU for HD 159062 B).  
Given that most barium stars have companions with orbital periods $\lesssim$10$^4$ days (e.g., \citealt{Mcclure:1980aa}),
that would naturally lead to the expectation that 12 Psc should be even \emph{more} enriched compared to HD 159062
because of more efficient accretion at periastron.  
This raises two open questions for the 12 Psc system:
Why was 12 Psc B not tidally circularized during the AGB phase?  
Why is 12 Psc unenriched in barium and $s$-process elements?
The chemical peculiarities of some barium stars have been attributed to winds from former AGB companions (now white dwarfs)
on orbits out to several thousand AU (\citealt{DeMello:1997ew}).
It remains unclear why some stars with white dwarf companions at moderate separations appear to have normal abundances while 
others show various patterns of enrichment.  The answer may involve convection and dissipation of material from the host star,
the amount of mass lost from the AGB companion, or perhaps a third evolved (and now engulfed) companion in the system.

12 Psc B and HD 159062 B join a small but growing list of directly imaged white dwarf 
companions orbiting main-sequence 
stars with measured dynamical masses.  
To our knowledge, only seven systems
with resolved white dwarf companions and precise dynamical mass measurements are 
known (including 12 Psc B and HD 159062 B; see compilation in Table~\ref{tab:wdcomps}).
These ``Sirius-like'' benchmark systems are valuable because they can be directly
characterized with photometry and spectroscopy---yielding an effective temperature, bolometric
luminosity, radius, and spectral classification---and the total system age 
and progenitor metallicity can be determined from the
host star.  These combined with a mass measurement provide fundamental 
tests of white dwarf mass-radius relations and cooling models (e.g., \citealt{Bond:2017hca}; \citealt{Bond:2017cx}; \citealt{Serenelli:2020aa}).
Follow-up spectroscopy and multi-wavelength photometry of 12 Psc B and HD 159062 B
are needed to better characterize these companions and carry out robust tests of cooling models.

\facility{Smith (Tull Spectrograph), HET (HRS), Keck:II (NIRC2)}

\acknowledgments

The authors are grateful to Michal Liu for the early NIRC2 imaging observations of 12 Psc in 2004 and 2005
as well as Keith Hawkins for helpful discussions about this system.
We thank Diane Paulson, Rob Wittenmyer, Erik Brugamyer, Caroline Caldwell, Paul Robertson, Kevin Gullikson, and Marshall Johnson for 
contributing to the Tull observations of 12 Psc presented in this study.
This work was supported by a NASA Keck PI Data Award, administered by the NASA Exoplanet Science Institute. Data presented herein were obtained at the W. M. Keck Observatory from telescope time allocated to the National Aeronautics and Space Administration through the agency's scientific partnership with the California Institute of Technology and the University of California. The Observatory was made possible by the generous financial support of the W. M. Keck Foundation.
This research has made use of the Keck Observatory Archive (KOA), which is operated by the W. M. Keck Observatory and the NASA Exoplanet Science Institute (NExScI), under contract with the National Aeronautics and Space Administration.
B.P.B. acknowledges support from the National Science Foundation grant AST-1909209.
The authors wish to recognize and acknowledge the very significant cultural role and reverence that the summit of Maunakea has always had within the indigenous Hawaiian community. We are most fortunate to have the opportunity to conduct observations from this mountain.

%\bibliographystyle{apj}
%\bibliography{oct20}

\end{document}

%% file: 12psc_rvs_tull_sample.tex
\begin{deluxetable}{lcc}
\renewcommand\arraystretch{0.9}
\tabletypesize{\small}
\setlength{ \tabcolsep } {.1cm}
\tablewidth{0pt}
\tablecolumns{3}
\tablecaption{Tull Spectrograph Relative Radial Velocities of 12 Psc\label{tab:12psc_rvs}}
\tablehead{
    \colhead{Date}  & \colhead{RV} & \colhead{$\sigma_{\mathrm{RV}}$} \\
    \colhead{(BJD)} & \colhead{(m s$^{-1}$)} & \colhead{(m s$^{-1}$)}
        }
\startdata
%========================================================
%           Date    &          RV    &      RV_err   \\
%========================================================
   2452115.95277    &      -97.23    &        3.69   \\
   2452145.89854    &      -98.83    &        4.20   \\
   2452219.66771    &      -86.14    &        5.03   \\
   2452473.90774    &      -59.28    &        4.44   \\
   2452494.83891    &      -75.32    &        6.33   \\
   2452540.79863    &      -65.29    &        5.13   \\
   2452598.71506    &      -52.06    &        4.50   \\
   2452896.84418    &      -48.27    &        5.57   \\
   2452931.83102    &      -73.12    &        4.71   \\
\multicolumn{3}{c}{$\cdots$} \\
\enddata
\tablecomments{Table 2 is published in its entirety in the machine-readable format.
      A portion is shown here for guidance regarding its form and content.}
\end{deluxetable}

%% file: hd159062_rvs_tull_sample.tex
\begin{deluxetable}{lcc}
\renewcommand\arraystretch{0.9}
\tabletypesize{\small}
\setlength{ \tabcolsep } {.1cm}
\tablewidth{0pt}
\tablecolumns{3}
\tablecaption{HRS Relative Radial Velocities of HD 159062\label{tab:hd159062_rvs}}
\tablehead{
    \colhead{Date}  & \colhead{RV} & \colhead{$\sigma_{\mathrm{RV}}$} \\
    \colhead{(BJD)} & \colhead{(m s$^{-1}$)} & \colhead{(m s$^{-1}$)}
        }
\startdata
%========================================================
%           Date    &          RV    &      RV_err   \\
%========================================================
   2454698.72895    &       34.67    &        3.77   \\
   2454715.68103    &       45.29    &        3.29   \\
   2454726.62674    &       45.12    &        3.95   \\
   2454728.63730    &       35.47    &        3.13   \\
   2454873.01630    &       42.84    &        2.98   \\
   2454889.98057    &       39.49    &        3.39   \\
   2454918.88920    &       47.88    &        3.21   \\
   2455059.74200    &       19.49    &        3.33   \\
   2455100.63545    &       16.31    &        3.76   \\
   2455261.94792    &       15.50    &        2.62   \\
\multicolumn{3}{c}{$\cdots$} \\
\enddata
\tablecomments{Table 3 is published in its entirety in the machine-readable format.
      A portion is shown here for guidance regarding its form and content.}
\end{deluxetable}

%% file: wd_12psc_mk_orbit_table.tex
\begin{deluxetable*}{lcccccc}
\renewcommand\arraystretch{0.9}
\tabletypesize{\small}
\setlength{ \tabcolsep } {.1cm} 
\tablewidth{0pt}
\tablecolumns{7}
\tablecaption{12 Psc B Orbit Fit Results \label{tab:12pscb_orbit}}
\tablehead{
    \colhead{Parameter} & \colhead{Prior} & \colhead{Best Fit} & \colhead{Median}  & \colhead{MAP\tablenotemark{a}}  & \colhead{68.3\% CI}  & \colhead{95.4\% CI}  
        }   
\startdata
\cutinhead{Fitted Parameters}
$M_1$ (M$_{\odot}$) & $\mathcal{N}$(1.1,0.2) & 1.10 & 1.10 & 1.07 & (0.91, 1.31) & (0.70, 1.51) \\
$M_2$ (M$_{\odot}$) & 1/$M_2$ & 0.594 & 0.605 & 0.594 & (0.583, 0.625) & (0.564, 0.648) \\
$a$ (AU) &  1/$a$ & 38.8 & 39.5 & 37.5 & (36.0, 42.3) & (35.4, 57.7) \\
$\sqrt{e} \sin \omega$ & $\mathcal{U}$(--1,1) & 0.70 & 0.65 & 0.68 & (0.30, 0.74) & (0.29, 0.96) \\
$\sqrt{e} \cos \omega$ & $\mathcal{U}$(--1,1) & --0.63 & --0.20 & 0.78 & (--0.68, 0.65) & (--0.67, 0.81) \\
$i$ ($^{\circ}$)& sin$i$ & 140 & 132 & 123 & (118, 151) & (108, 157) \\
$\Omega$ ($^{\circ}$) & $\mathcal{U}$(--180, 180) & --4.77 & --42.2 & --142 & (--145, --7.2) & (--145, 13.3) \\
$\lambda_{\mathrm{ref}}$ ($^{\circ}$)\tablenotemark{b} & $\mathcal{U}$(--180, 180) & 17.9 & --23.3 & 34.5 & (--125, 42.4) & (--163, 45.9) \\
\cutinhead{Derived Parameters}
$e$ &  $\cdots$  &0.89 & 0.84 & 0.89 & (0.76, 0.92) & (0.27, 0.93) \\
$\omega$ ($^{\circ}$)& $\cdots$  & 132 & 104 & 25.5 & (33.9, 138) & (22.0, 138) \\
$P$ (yr) &  $\cdots$  &186 & 193 & 175 & (155, 218) & (145, 348) \\
$\tau$ (yr)\tablenotemark{c} &  $\cdots$  &2069 & 2074 & 2070 & (2060, 2080) & (2060, 2180) \\
$d_p$ (AU) &  $\cdots$  &4.5 & 6.4 & 3.5 & (2.7, 10.2) & (2.4, 40.2) \\
$d_a$ (AU) &  $\cdots$  &73.2 & 71.2 & 68.5 & (66.9, 73.9) & (66.2, 85.1) \\
\enddata
\tablenotetext{a}{Maximum a posteriori probability.}
\tablenotetext{b}{Mean longitude at the reference epoch, 2455197.5 JD.}
\tablenotetext{c}{Time of periastron, 2455197.5 JD -- $P$($\lambda_{\mathrm{ref}}$ -- $\omega$)/(2$\pi$).}
\end{deluxetable*}

%% file: wd_hd159062_mk_orbit_table.tex
\begin{deluxetable*}{lcccccc}
\renewcommand\arraystretch{0.9}
\tabletypesize{\small}
\setlength{ \tabcolsep } {.1cm} 
\tablewidth{0pt}
\tablecolumns{7}
\tablecaption{HD 159062 B Orbit Fit Results \label{tab:hd159062b_orbit}}
\tablehead{
    \colhead{Parameter} & \colhead{Prior} & \colhead{Best Fit} & \colhead{Median}  & \colhead{MAP\tablenotemark{a}}  & \colhead{68.3\% CI}  & \colhead{95.4\% CI}  
        }   
\startdata
\cutinhead{Fitted Parameters}
$M_1$ (M$_{\odot}$) & $\mathcal{N}$(0.8,0.2) & 0.896 & 0.799 & 0.770 & (0.62, 0.97) & (0.46, 1.16) \\
$M_2$ (M$_{\odot}$) & 1/$M_2$ & 0.597 & 0.609 & 0.610 & (0.599, 0.619) & (0.588, 0.630) \\
$a$ (AU) &  1/$a$ & 62.3 & 59.9 & 62.5 & (52.7, 65.0) & (42.7, 71.0) \\
$\sqrt{e} \sin \omega$ & $\mathcal{U}$(--1,1) & 0.021 & --0.14 & --0.20 & (--0.30, --0.02) & (--0.35, 0.18) \\
$\sqrt{e} \cos \omega$ & $\mathcal{U}$(--1,1) & 0.11 & 0.18 & 0.28 & (--0.07, 0.48) & (--0.36, 0.64) \\
$i$ ($^{\circ}$)& sin$i$ & 64.6 & 61.9 & 62.3 & (59.5, 64.9) & (54.1, 67.1) \\
$\Omega$ ($^{\circ}$) & $\mathcal{U}$(--180, 180) & 133 & 134 & 134 & (132, 140) & (130, 141) \\
$\lambda_{\mathrm{ref}}$ ($^{\circ}$)\tablenotemark{b} & $\mathcal{U}$(--180, 180) & 146 & 147 & 148 & (141, 155) & (125, 161) \\
\cutinhead{Derived Parameters}
$e$ &  $\cdots$  &0.013 & 0.092 & 0.010 & (0.00, 0.15) & (0.00, 0.40) \\
$\omega$ ($^{\circ}$)& $\cdots$  & 10.7 & 139 & 156 & (103, 180) & (8.21, 180) \\
$P$ (yr) &  $\cdots$  &402 & 387 & 407 & (314, 457) & (230, 533) \\
$\tau$ (yr)\tablenotemark{c} &  $\cdots$  &1858 & 2000 & 2025 & (1950, 2050) & (1840, 2050) \\
$d_p$ (AU) &  $\cdots$  &61 & 56 & 63 & (49, 64) & (28, 64) \\
$d_a$ (AU) &  $\cdots$  &63 & 64 & 64 & (62, 65) & (62, 78) \\
\enddata
\tablenotetext{a}{Maximum a posteriori probability.}
\tablenotetext{b}{Mean longitude at the reference epoch, 2455197.5 JD.}
\tablenotetext{c}{Time of periastron, 2455197.5 JD -- $P$($\lambda_{\mathrm{ref}}$ -- $\omega$)/(2$\pi$).}
\end{deluxetable*}